\documentclass[12pt]{article}
\begin{document}
\newcommand{\beq}{\begin{equation}}
\newcommand{\eeq}{\end{equation}}
\newcommand{\beqa}{\begin{eqnarray}}
\newcommand{\eeqa}{\end{eqnarray}}
\newcommand{\beqar}{\begin{eqnarray*}}
\newcommand{\eeqar}{\end{eqnarray*}}
\newcommand{\al}{\alpha}
\newcommand{\be}{\beta}
\newcommand{\del}{\delta}
\newcommand{\D}{\Delta}
\newcommand{\eps}{\epsilon}
\newcommand{\ga}{\gamma}
\newcommand{\Ga}{\Gamma}
\newcommand{\ka}{\kappa}
\newcommand{\nn}{\nonumber}
\newcommand{\inn}{\!\cdot\!}
\newcommand{\h}{\eta}
\newcommand{\ii}{\iota}
\newcommand{\kk}{\varphi}
\newcommand\F{{}_3F_2}
\newcommand{\la}{\lambda}
\newcommand{\La}{\Lambda}
\newcommand{\na}{\prt}
\newcommand{\Om}{\Omega}
\newcommand{\om}{\omega}
\newcommand{\p}{\phi}
\newcommand{\sig}{\sigma}
\renewcommand{\t}{\theta}
\newcommand{\z}{\zeta}
\newcommand{\ssc}{\scriptscriptstyle}
\newcommand{\eg}{{\it e.g.,}\ }
\newcommand{\ie}{{\it i.e.,}\ }
\newcommand{\labell}[1]{\label{#1}} 
\newcommand{\reef}[1]{(\ref{#1})}
\newcommand\prt{\partial}
\newcommand\veps{\varepsilon}
\newcommand{\pol}{\varepsilon}
\newcommand\vp{\varphi}
\newcommand\ls{\ell_s}
\newcommand\cF{{\cal F}}
\newcommand\cA{{\cal A}}
\newcommand\cS{{\cal S}}
\newcommand\cT{{\cal T}}
\newcommand\cV{{\cal V}}
\newcommand\cL{{\cal L}}
\newcommand\cM{{\cal M}}
\newcommand\cN{{\cal N}}
\newcommand\cD{{\cal D}}
\newcommand\cH{{\cal H}}
\newcommand\cI{{\cal I}}
\newcommand\cJ{{\cal J}}
\newcommand\cl{{\iota}}
\newcommand\cP{{\cal P}}
\newcommand\cQ{{\cal Q}}
\newcommand\cg{{\it g}}
\newcommand\cR{{\cal R}}
\newcommand\cB{{\cal B}}
\newcommand\cO{{\cal O}}
\newcommand\tcO{{\tilde {{\cal O}}}}
\newcommand\bg{\bar{g}}
\newcommand\bb{\bar{b}}
\newcommand\bH{\bar{H}}
\newcommand\bX{\bar{X}}
\newcommand\bK{\bar{K}}
\newcommand\bA{\bar{A}}
\newcommand\bZ{\bar{Z}}
\newcommand\bxi{\bar{\xi}}
\newcommand\bphi{\bar{\phi}}
\newcommand\bpsi{\bar{\psi}}
\newcommand\bprt{\bar{\prt}}
\newcommand\bet{\bar{\eta}}
\newcommand\btau{\bar{\tau}}
\newcommand\bc{\bar{c}}
\newcommand\hF{\hat{F}}
\newcommand\hA{\hat{A}}
\newcommand\hT{\hat{T}}
\newcommand\htau{\hat{\tau}}
\newcommand\hD{\hat{D}}
\newcommand\hf{\hat{f}}
\newcommand\hg{\hat{g}}
\newcommand\hp{\hat{\phi}}
\newcommand\hi{\hat{i}}
\newcommand\ha{\hat{a}}
\newcommand\hb{\hat{b}}
\newcommand\hQ{\hat{Q}}
\newcommand\hP{\hat{\Phi}}
\newcommand\hS{\hat{S}}
\newcommand\hX{\hat{X}}
\newcommand\tL{\tilde{\cal L}}
\newcommand\hL{\hat{\cal L}}
\newcommand\tG{{\widetilde G}}
\newcommand\tg{{\widetilde g}}
\newcommand\tphi{{\widetilde \phi}}
\newcommand\tPhi{{\widetilde \Phi}}
\newcommand\te{{\tilde e}}
\newcommand\tk{{\tilde k}}
\newcommand\tf{{\tilde f}}
\newcommand\ta{{\tilde a}}
\newcommand\tb{{\tilde b}}
\newcommand\ti{{\tilde i}}
\newcommand\tj{{\tilde j}}
\newcommand\tR{{\tilde R}}
\newcommand\teta{{\tilde \eta}}
\newcommand\tF{{\widetilde F}}
\newcommand\tK{{\widetilde K}}
\newcommand\tE{{\widetilde E}}
\newcommand\tpsi{{\tilde \psi}}
\newcommand\tX{{\widetilde X}}
\newcommand\tD{{\widetilde D}}
\newcommand\tO{{\widetilde O}}
\newcommand\tS{{\tilde S}}
\newcommand\tB{{\widetilde B}}
\newcommand\tA{{\widetilde A}}
\newcommand\tT{{\widetilde T}}
\newcommand\tC{{\widetilde C}}
\newcommand\tV{{\widetilde V}}
\newcommand\thF{{\widetilde {\hat {F}}}}
\newcommand\Tr{{\rm Tr}}
\newcommand\tr{{\rm tr}}
\newcommand\STr{{\rm STr}}
\newcommand\hR{\hat{R}}
\newcommand\M[2]{M^{#1}{}_{#2}}

\newcommand\bS{\textbf{ S}}
\newcommand\bI{\textbf{ I}}
\newcommand\bJ{\textbf{ J}}

\begin{titlepage}
\begin{center}

\vskip 2 cm
{\LARGE \bf     O-plane  couplings at order $\alpha'^2$:  \\ \vskip 0.5 cm
one R-R  field strength
 }\\
\vskip 1.25 cm
  Mahboube Mashhadi\footnote{se.mashhadi@mail.um.ac.ir} and   Mohammad R. Garousi\footnote{garousi@um.ac.ir} 
 
\vskip 1 cm
{{\it Department of Physics, Faculty of Science, Ferdowsi University of Mashhad\\}}
\vskip .1 cm
 \end{center}

\begin{abstract}

   It is known that the   anomalous   Chern-Simons (CS) coupling  of O$_p$-plane   is not consistent with the  T-duality transformations. Compatibility of this coupling with the  T-duality  requires  
   the  inclusion of   couplings   involving  one R-R field strength.  In this paper we find such couplings at order $\alpha'^2$. 
   
By requiring  the R-R and NS-NS gauge invariances, we first find all independent couplings at order $\alpha'^2$. 
There are $1,\, 6,\,28,\,20,\, 19,\, 2$ couplings 
corresponding to    the R-R field strengths $F^{(p-4)}$, $\,F^{(p-2)}$, $\,F^{(p)}$, $\,F^{(p+2)}$, $\,F^{(p+4)}$  and $F^{(p+6)}$, respectively. We then impose the T-duality constraint on these couplings and on the CS coupling $C^{(p-3)}\wedge R\wedge R$  at order $\alpha'^2$   to fix their corresponding     coefficients. The T-duality constraint  fixes  all    coefficients in terms of the CS coefficient.  
    They  are   fully consistent with the partial  couplings that have been already found  in the literature by  the S-matrix method.  
\end{abstract}
\end{titlepage}

\section{Introduction}

The best candidate for quantum gravity is the superstring theory in which the graviton    appears as  a specific mode of a relativistic superstring  at weak coupling \cite{Polchinski:1996na,Becker:2007zj}. Superstring  has massless and infinite tower of massive states which appear in the low energy  effective action,   as higher derivative corrections to the supergravity. Study of these higher derivative corrections are important because they signal the stringy nature of the quantum gravity. 

One of the most exciting discoveries in perturbative  string theory is  the  T-duality which has been observed  first  in the spectrum  of string  when one compactifies theory  on a circle \cite{Giveon:1994fu,Alvarez:1994dn}. This symmetry may be used to construct the effective action of string theory including its higher derivative corrections, in  the  Double Field Theory formalism  in which the T-duality transformations are the standard $O(D,D)$ transformations whereas the  gauge transformations are non-standard  \cite{Siegel:1993xq,Hull:2009mi,Hohm:2010jy}. It has been also speculated  that the invariance of the effective actions of string theory and   its non-perturbative objects, \ie D-branes and O-planes,  under    the standard  gauge transformations  and    non-standard   T-duality transformations   may be used as  a constraint to construct the   effective actions \cite{Garousi:2017fbe}. In this approach,  
one first constructs the most general   gauge invariant and independent couplings at a given order of $\alpha'$ with arbitrary parameters. Then the parameters   may  be fixed in  the  string theory  by imposing the T-duality symmetry on the couplings. That is, one reduces the couplings  on   a circle and requires them to be consistent with the T-duality transformations which are  the standard Buscher rules  \cite{Buscher:1987sk,Buscher:1987qj} plus their  $\alpha'$-corrections  \cite{Tseytlin:1991wr,Bergshoeff:1995cg,Kaloper:1997ux,Garousi:2019wgz}.       Using this approach, the effective action of the bosonic string theory at order $\alpha'$ and $\alpha'^2$  have been found in \cite{Garousi:2019wgz,Garousi:2019mca}. It has been shown in \cite{Garousi:2019xlf,Garousi:2019jbq} that  the leading order effective action of type II superstring theories,  including  the Gibbons-Hawking-York boundary term \cite{York:1972sj,Gibbons:1976ue},    can also be  rederived  by the T-duality constraint. The   couplings involving metric and dilaton  in both  heterotic string and in  superstring theories  at order $\alpha'^3$ have been also rederived by the T-duality constraint in \cite{Razaghian:2018svg}. There are many other approaches for constructing the effective actions including   the S-matrix   approach \cite{Scherk:1974mc,Yoneya:1974jg}, the sigma-model approach \cite{Callan:1985ia,Fradkin:1984pq,Fradkin:1985fq},  and the supersymmetry approach \cite{Gates:1986dm,Gates:1985wh,Bergshoeff:1986wc,Bergshoeff:1989de}  

The T-duality   approach for constructing   the effective action of D$_p$-brane (O$_p$-plane)   is  such that one first writes all  gauge invariant and independent D$_p$-brane (O$_p$-plane) world-volume couplings at a specific order of $\alpha'$ with some unknown $p$-independent   coefficients. Then one reduces the world-volume theory on the circle. There are two possibilities for the killing coordinate. Either it is along   or  orthogonal to the brane. The reduction of the world-volume theory when the killing coordinate is along the brane (the world-volume reduction),  is different from the reduction of the world-volume theory when the killing coordinate is orthogonal to the brane (the transverse reduction). However, the T-duality transformation of  the world-volume reduction of D$_p$-brane  (O$_p$-plane)   should be  the same as the transverse reduction of the   D$_{p-1}$-brane (O$_{p-1}$-plane) theory, up to some total derivative terms which have no physical effects  for closed spacetime manifold \cite{Garousi:2017fbe}. 

Since  O$_p$-planes are at the fixed points of spactime, \ie $X^i=0$,  some world-volume couplings are forbidden by orientifold projection \cite{Polchinski:1996na}. The O$_p$-plane effective action has no   open string couplings, no  couplings that have odd number of transverse indices on metric and dilaton and their corresponding derivatives, and no couplings that have even number of transverse indices on B-field and its corresponding derivatives \cite{Polchinski:1996na}.  These O-plane conditions make  the study of the  O-plane couplings to be much easier than  the D-brane couplings.  The T-duality  constraint has been used in \cite{Robbins:2014ara,Garousi:2014oya} to find the effective action of O$_p$-planes of type II superstring  at order $\alpha'^2$ for NS-NS fields. In this paper, we are interested in applying the T-duality constraint on the effective action of O$_p$-plane when there is one R-R field strength. 

   The O$_p$-plane  CS action at the leading order of $\alpha'$ is  given as \cite{Polchinski:1996na} 
\beqa
S_{CS}^{(0)}&=&T_p\int_{M^{p+1}} C\labell{BC}
\eeqa
where $C=\sum_{n=0}^{8}C^{(n)}$ is the R-R potential and $T_p$ is the O$_p$-plane tension. It is invariant under the R-R gauge transformation
\beqa
\delta C&=&d\Lambda+H\Lambda\labell{dc}
\eeqa
 where $\Lambda=\sum_{n=0}^7\Lambda^{(n)}$ and $H=dB$. Note that the last term  in $\delta C$ is zero for O$_p$-plane  when all indices of the R-R potential are world-volume, as in \reef{BC}.  The curvature corrections to D$_p$-brane action has been found 
 by requiring that the chiral anomaly on the world-volume of intersecting D-branes  cancels with the anomalous variation of the CS action \cite{Green:1996dd,Cheung:1997az,Minasian:1997mm}. The corresponding corrections for O$_p$-plane  has been found in \cite{Morales:1998ux} to be 
 \beqa
S_{CS}&=&T_{p}\int_{M^{p+1}}C\sqrt{\frac{{\cal L}(\pi^2\alpha'R_T)}{{\cal L}(\pi^2\alpha'R_N)}}\labell{CS}
\eeqa
where   ${\cal L}(R_{T,N})$ is the Hirzebruch polynomials of the tangent and normal bundle curvatures respectively,
\beqa
\sqrt{\frac{{\cal L}(\pi^2\alpha'R_T)}{{\cal L}(\pi^2\alpha'R_N)}}&=&1-\frac{\pi^2\alpha'^2}{48}(\tr R_T^2-\tr R_N^2)+\cdots \labell{roof}
\eeqa
where  $R_{T,N}$ are the   curvature 2-forms of the tangent and normal bundles respectively. The corresponding curvature corrections to the CS action of D$_p$-brane is the same as \reef{roof} in which ${\cal L}(R/4)$ is replaced by the A-roof genus ${\cal A}(R)$ which produces up to a factor of $-2$,  the same curvature corrections at order $\alpha'^2$. However, the curvature corrections at higher orders of order $\alpha'$ are not the same in both cases.

The action \reef{roof} at order $\alpha'^2$ in component form is\footnote{Our index convention is that $A,B,\cdots$ are 10-dimensional bulk indices, $\mu,\nu,\cdots$ are 9-dimensional base  indices, $y$ is killing index, $a,b,\cdots$ and $a_0,\cdots, a_p$ are world-volume indices and $i,j,\cdots$ are transverse space indices.}
\beqa
S_{CS}^{(2)}&=&-\frac{T_p\pi^2\alpha'^2}{48}\int d^{p+1}x\epsilon^{a_0\cdots a_p}\frac{1}{4(p-3)!}{ C}^{(p-3)}_{a_4\cdots a_{p}}\bigg[R_{a_{0}a_{1}}{}^{ab}R_{a_{2}a_3\,ab}-R_{a_{0}a_{1}}{}^{ij}R_{a_{2}a_3\,ij}\bigg]\labell{Tf2}
\eeqa
The above  couplings have been confirmed by the S-matrix element calculations in \cite{Craps:1998fn,Morales:1998ux,Stefanski:1998yx}. Using the cyclic symmetry of the Riemann curvature, one can verify that the above diffeomorphism invariant action is also  invariant under R-R gauge transformation.  As it has been argued in  \cite{Myers:1999ps,Becker:2009rn}, the above couplings, however, are not consistent with the T-duality transformations. 

On the other hand, there are  many other gauge invariant couplings at this order which can not be found by the anomaly analysis.  The R-R gauge symmetry requires all such couplings to be in terms of the nonlinear R-R field strength, \ie
\beqa
F^{(n)}&=& dC^{(n-1)}+H\wedge C^{(n-3)}\labell{Fn}
\eeqa
which is invariant under the R-R gauge transformation \reef{dc}.  Some of these couplings  involving one R-R  field strength $F^{(p-2)}$ and two NS-NS fields have been found for D-brane in \cite{  Becker:2010ij,Garousi:2010rn,Garousi:2011ut,Becker:2011ar}  by linear T-duality and by the disk-level S-matrix calculations. The complete couplings involving one R-R  field strength $F^{(p)}$, $F^{(p+2)}$ or $F^{(p+4)}$  and one NS-NS field have been found in  \cite{Garousi:2010ki}  by the S-matrix method and have been shown that they are invariant under the linear T-duality.
However,  these couplings are not invariant under the full nonlinear T-duality either.  Hence, the T-duality  of the CS coupling \reef{Tf2}   may require 
  adding couplings involving one R-R field strength and an arbitrary number of NS-NS fields at order $\alpha'^2$ in which we are interested in this paper.
 
 An outline of the paper is as follows: In section 2, we find the minimal  gauge invariant couplings involving one R-R field strength. We use the Bianchi identities, total derivative terms and  $\epsilon^{a_0\cdots a_p}$-tensor identities to find the minimum number of  gauge invariant couplings. We find there are  $1,\, 6,\,28,\,20,\, 19,\, 2$  such couplings corresponding to    the R-R field strengths $F^{(p-4)}$, $\,F^{(p-2)}$,$F^{(p)}$, $F^{(p+2)}$, $F^{(p+4)}$, $F^{(p+6)}$, respectively.  We then reduce them on a circle in section 3 to impose the T-duality constraint on them. 
 
  An appropriate    method for reducing a gauge invariant coupling to 9-dimensional base space has been presented in \cite{Garousi:2019mca}. In this method one keeps the $U(1)\times U(1)$ gauge invariant part in the reduction of the Riemann curvature and other components of a given coupling and removes all other terms.   In section 3, we extend this method   for the reduction of the couplings involving   R-R fields as well, \ie    we find the $U(1)\times U(1)$ gauge invariant part of the reduction of R-R field strength and its first derivatives. In section 4, we impose the T-duality constraint on the independent gauge invariant couplings to fix  their  parameters. That is, we use the Bianchi identities, total derivative terms and  $\epsilon^{a_0\cdots a_{p-1}}$-tensor identities in the base space to write the T-duality constraint in terms of  independent structures, and then solve them. In this section, we show that the T-duality can fix  all parameters of the gauge invariant couplings in terms of an overall factor, and they are consistent with the partial  couplings that have been already found in the literature by the S-matrix method.  
  In section 5, we present the final form of the gauge invariant  couplings and briefly discuss our results.

\section{Minimal gauge invariant   couplings}

In this section we would like to find minimum number of gauge invariant  couplings on the world-volume of O$_p$-plane involving one R-R field strength   and an arbitrary number of  NS-NS fields at order $\alpha'^2$, \ie
\beqa
S_n&=&-\frac{T_p\pi^2\alpha'^2}{48}\int d^{p+1}x\,{\cal L}^n\labell{Gen}
\eeqa
where ${\cal L}^n$  is the Lagrangian which includes the minimum number of gauge invariant  couplings involving one R-R field strength $F^{(n)}$. As it has been argued in \cite{Robbins:2014ara},  since we are interested in O$_p$-plane as a probe, it does not have back reaction on the spacetime. As a result, the massless closed string fields must satisfy the bulk equations of motion at order $\alpha'^0$. Using the equations of motion, one can rewrite the terms in the world-volume theory which have  contraction of two transverse indices, \eg $\nabla_i\nabla^i\Phi$, or $R_{iA}{}^i{}_B$ in terms of contraction of two world-volume indices, \eg  $\nabla_a\nabla^a\Phi$, or $R_{aA}{}^a{}_B$. This indicates that the former couplings are not independent. The O-plane couplings should also satisfy the orientifold projection.

 The couplings involving the Riemann curvature and its derivative and the couplings involving derivatives of H and derivatives of R-R field strength satisfy the following Bianchi identities
\beqa
R_{A[BCD]}&=&0\nn\\
\nabla_{[A}R_{BC]DE}&=&0\nn\\
dH&=&0\nn\\
dF^{(n)}+H\wedge F^{(n-2)}&=&0\labell{Bian}
\eeqa
Moreover, the couplings involving the commutator of two covariant derivatives of a tensor are  not independent of the couplings involving the contraction of this tensor with the  Riemann curvature, \ie  
\beqa
[\nabla, \nabla ]{\cal O}&=&R{\cal O}\labell{DD}
\eeqa
This indicates that if one considers all gauge invariant couplings at a given order of $\alpha'$, then only one ordering of the covariant derivatives is needed to be considered.  

Using the symmetries of $\epsilon^{a_0\cdots a_p}$, the R-R field strength $F^{(n)}$, $H$ and the Riemann curvature, one can easily verify  that it is impossible to have non-zero contractions of one $F^{(n)}$ and some $R,\,H,\, \nabla\Phi$ at order $\alpha'^2$ for $n<p-4$ and $ n>p+6$. Moreover,  the parity of the  coupling \reef{Tf2} indicates that   the couplings of the  R-R field strength $F^{(p-2)}$ are non-zero when there are even number of B-field. The consistency with linear T-duality then indicates that the couplings of  the  R-R field strength $F^{(p-4)}$,    $F^{(p)}$ and $F^{(p+4)}$ are non-zero when there are odd number of B-field, and the couplings of  the  R-R field strength $F^{(p+2)}$,  and   $F^{(p+6)}$   are non-zero when there are even number of B-field. There are similar parity selection rule  for the corresponding S-matrix elements \cite{Velni:2013jha}. For $n=p-4$ there is only one non-zero independent coupling\footnote{We have used the package ''xAct" \cite{Nutma:2013zea} for performing the calculations in this paper.}, \ie
\beqa
\cL^{p-4}&=&\epsilon^{a_0\cdots a_p}\Big[\frac{a}{(p-5)!}F_{ia_6\cdots a_p}H^i{}_{a_0a_1}H_{ja_2a_3}H^j{}_{a_4a_5}\Big]\labell{LP-4}
\eeqa
where we have used the O-plane conditions that there is no $H$ term with even number of transverse indices. In above equation, the transverse indices are raised by   the tensor  $\bot^{ij}=G^{ij}$ (see next section for the definition of tensor $\bot$), and coefficient  $a$ is an arbitrary parameter at this point. This parameter may be fixed by studying the  $RP^2$-level  S-matrix element of one R-R   and three NS-NS vertex operators   which is a very lengthy calculation. We expect this parameter to be fixed by  the T-duality constraint. 

There is no derivative on the R-R field strength and on the B-field strength  in the above coupling. Hence, there is no Bianchi identity involved here. Since there is only one term, there would be no $\epsilon$-tensor identity either. Moreover, there is no total derivative term here. This is not the case for $n>p-4$ cases. Let us discuss each of the cases $n=p-2$, $n=p$, $n=p+2$, $n=p+4$ and $n=p+6$ separately. 

\subsection{$n=p-2$ case}

To find all  gauge invariant and independent couplings corresponding to  one  R-R field strength $F^{(p-2)}$,  we first consider all contractions of one $\epsilon^{a_0\cdots a_p}$, one $F$, $\nabla F$ or $\nabla\nabla F$, even number of $H$ and $ \nabla H$, and any number of $\nabla\Phi$, $\nabla\nabla\Phi$, $\nabla\nabla\nabla\Phi$, $R,\,\nabla R$ at four-derivative order. Because of the relation \reef{DD}, we consider only one ordering  of the covariant derivatives. We then remove the  forbidden couplings for O-plane, and remove the couplings in which two transverse indices in a term contracted, \ie we impose   the equations of motion. We call the remaining terms, with coefficients $b'_1, b'_2, \cdots $,  the Lagrangian  $L^{p-2}$. Not all terms in this Lagrangian, however,  are  independent. Some of them are related by total derivative terms, by Bianchi identity and by $\epsilon$-tensor identity. 

To remove the total derivative redundancy, we write all total derivative terms at order $\alpha'^2$ which involve  the  R-R field strength   $F^{(p-2)}$. To this end we first write   all contractions of one $\epsilon^{a_0\cdots a_p}$, one $F$, $\nabla F$,    even number of $H$ and $ \nabla H$, and any number of $\nabla\Phi$, $\nabla\nabla\Phi$,   $R$ at three-derivative order.  Then we remove the forbidden couplings and impose  the equations of motion. We call the remaining terms, with arbitrary coefficients,  the vector $I^{p-2}_a$. The total derivative terms are then
\beqa
J^{p-2}&=&\int d^{p+1}x \,\tg^{ab}\nabla_a I^{p-2}_b
\eeqa
where $\tg^{ab}=G^{ab}$ is inverse of the pull-back metric (see next section for the definition of the pull-back metric). Adding the total derivative terms to $L^{p-2}$, one finds the same Lagrangian     but with different parameters $b_1, b_2, \cdots$. We call the new Lagrangian  ${\cal L}^{p-2}$. Hence 
\beqa
\Delta^{p-2}-J^{p-2}&=&0\labell{DL}
\eeqa
where $\Delta^{p-2}={\cal L}^{p-2}-L^{p-2}$ is the same as $L^{p-2}$ but with coefficients $\delta b_1,\delta b_2,\cdots$ where $\delta b_i= b_i-b'_i$. Solving the above equation, one would find some linear  relations between  only $\delta b_1,\delta b_2,\cdots$ which indicate how the couplings are related among themselves by the total derivative terms. The above equation would also give some relation between the coefficients of the total derivative terms and $\delta b_1,\delta b_2,\cdots$ in which we are not interested.

However, to solve the above equation one has to impose the Bianchi identity and $\epsilon$-tensor identities.  To impose the Riemann curvature and $H$-field Bianchi identities \reef{Bian}, one may contract the term on the left-hand side of each Bianchi identity with appropriate couplings to produce  terms at order $\alpha'^2$. The coefficients of these terms are also arbitrary. Adding these terms to the equation \reef{DL}, then one could solve the equation to find the linear relations between   only $\delta b_1,\delta b_2,\cdots$. This method has been used in \cite{Robbins:2014ara} to find the independent couplings involving only the NS-NS fields.  Alternatively, to impose the  Riemann curvature Bianchi identities, one may rewrite the terms in \reef{DL} in  the local frame in which the first derivative of metric is zero. Similarly, to impose the H-field Bianchi identity, one may  rewrite the terms in \reef{DL} which have derivatives of $H$ in terms of B-field potential, \ie $H=dB$.  The last Bianchi identity in \reef{Bian} relates the couplings involving derivative of $F^{(p-2)}$ to  themselves and to the couplings involving $F^{(p-4)}$. However, the independent couplings involving  $F^{(p-4)}$ have been already fixed in \reef{LP-4}. Hence, the last Bianchi identity in \reef{Bian} should relate only the couplings  involving $F^{(p-2)}$, \ie one should impose the identity $dF^{(p-2)}=0$. To impose this identity  on the couplings in \reef{DL} as well, one may  rewrite the terms involving the derivatives of the R-R field strength $F^{(p-2)}$ in terms of the R-R potential, \ie   $F^{(p-2)}=dC^{(p-3)}$. In this way,  all Bianchi identities satisfy automatically \cite{Garousi:2019cdn}. We find that this latter approach is easier to impose the Bianchi identities by computer. Moreover, in this approach one does not need to introduce a large number of arbitrary parameters to include the Bianchi identities to the equation \reef{DL}. However, in this approach the gauge invariant equation \reef{DL} is written in terms of non-gauge invariant couplings.  In this paper we use this approach for imposing the Bianchi identities.

After imposing the Bianchi identities, the non-gauge invariant couplings are not yet independent. To rewrite them  in terms of independent couplings, one has to use the fact that  the number of world-volume indices in each coupling must be the same as the world-volume  indices of $\epsilon^{a_0\cdots a_p}$. It has been observed in \cite{Garousi:2010ki} that imposing this constraint, one may find some relations between couplings involving  $\epsilon^{a_0\cdots a_p}$. Some of these $\epsilon$-tensor identities  for the    simple case  of two-field couplings, have been found in \cite{Garousi:2010ki}. To impose this constraint on the couplings in  \reef{DL} as well, we write the non-gauge invariant couplings   explicitly in terms of the   values that each world-volume  index can take, \eg $a_0=0,1,2,\cdots,p$. It is easy to perform this step by computer using the  ''xAct" package \cite{Nutma:2013zea}. 

Using the above  steps, one can rewrite the different gauge invariant couplings on the left-hand side of \reef{DL} in terms of independent but non-gauge invariant couplings. The solution to the equation \reef{DL} then has two parts. One part is  relations between only $\delta b_i$'s, and the other part is a relation between the coefficients of the total derivative terms and $\delta b_i$'s in which we are not interested. The number of relations in the first part gives the minimum number  of gauge invariant  couplings in ${\cal L}^{p-2}$. To write the independent couplings in a specific scheme, one must set some of the coefficients in  $L^{p-2}$ to zero. However, after replacing the non-zero terms in \reef{DL}, the number of relations between only $\delta b_i$'s should not be changed. In the present case this number is 6. We set the coefficients of the terms that have world-volume derivative on the R-R field strength, to be zero. After setting this coefficients to zero, there are still 6 relations between  $\delta b_i$'s.  This means we are  allowed  to  remove these terms.  We choose some other coefficients to zero such that the remaining coefficients satisfy the 6 relations $\delta b_i=0$.    In this way one can find the minimum number of gauge invariant couplings.  One particular choice for the 6  couplings is the following\footnote{If one does not use the $\epsilon$-tensor identities, then one would find 10 independent couplings.}:
\beqa
{\cal L}^{(p-2)}&\!\!\!\!\!=\!\!\!\!\!\!&\epsilon^{a_{0}...a_{p}}\Big[\frac{b_1}{(p-3)!}\,  \nabla_{i}F_{ja_{4}...a_{p}}\, H^{i}\,_{a_{0}a_{1}}\, H^{j}\,_{a_{2}a_{3}}
+\frac{b_2}{(p-2)!}\, F_{a_{3}...a_{p}}\, \nabla^{a}H_{iaa_{0}}\, H^{i}\,_{a_{1}a_{2}}\nn\\
&& 
+\frac{b_4}{(p-2)!}\, F_{a_{3}...a_{p}}\, \nabla_{a_{0}}H_{iaa_{1}}\, H^{ia}\,_{a_{2}}+\frac{b_5}{(p-2)!}\, F_{a_{3}...a_{p}}\, \nabla_{a}H_{ia_{0}a_{1}}\, H^{ia}\,_{a_{2}}\nn\\
&&+\frac{b_7}{(p-4)!}\, F_{ij a_{5}...a_{p}}\, \nabla_{a_{0}}H^{i}\,_{a_{1}a_{2}}\, H^{j}\,_{a_{3}a_{4}}
+\frac{b_9}{(p-2)!}\, F_{a_{3}...a_{p}}\, H_{iaa_{0}}\, H^{i}\,_{a_{1}a_{2}}\, \nabla^{a} \Phi
\Big]\labell{Lp-2}
\eeqa
where  the world-volume indices are raised by the first fundamental form  $\tG^{ab}=G^{ab}$ (see next section for the definition of the first fundamental form), and  the   $b$'s are arbitrary coefficients.  These coefficients do not depend on $p$. In fact the $p$-dependence of the couplings has been written  explicitly by $1/n!$ where $n$ is the number of indices of the R-R field strength that are contracted with $\epsilon^{a_0\cdots a_p}$.  These couplings are consistent with the linear T-duality for the special case that   the world-volume killing index of  $\epsilon^{a_0\cdots a_{p}}$  contracts with the R-R field strength. That is, 
\beqa
\frac{1}{(p+1-m)!}\epsilon^{a_0\cdots a_ma_{m+1}\cdots a_{p}}F_{\cdots a_{m}a_{m+1}\cdots a_p}(\cdots)&=&\frac{1}{(p-m)!}\epsilon^{a_0\cdots a_{p-1}y}F_{\cdots  a_{m}a_{m+1}\cdots a_{p-1}y}(\cdots)+\cdots\nn\\
 &\rightarrow & \frac{1}{(p-m)!}\epsilon^{a_0\cdots a_{p-1}}F_{\cdots  a_{m}a_{m+1}\cdots a_{p-1}}(\cdots)+\cdots\nn
\eeqa
where the dots before the index $a_m$ in the R-R field strength are the world-volume or transverse indices  that contract with other parts of the coupling, \ie contract with $(\cdots)$. In the first line we assume one of the world-volume indices is the killing index $y$, and  in the second line we have used the linear T-duality transformation for the linearised R-R field strength, \ie $F^{(n)}_{\cdots y}=F^{(n-1)}_{\cdots}$, and the identity $\epsilon^{a_0\cdots a_{p-1}y}=\epsilon^{a_0\cdots a_{p-1}}$. The  couplings \reef{Lp-2} for arbitrary coefficients, however, are not  consistent with the linear T-duality when the killing index is not carried by the R-R field strength. We are interested in constricting  these  coefficients and the coefficients of other R-R field strengths that we will find in the subsequent subsections, by  requiring  the couplings to be consistent with nonlinear T-duality.

There is no term in \reef{Lp-2} which involves only one NS-NS field. This indicates that the $RP^2$-level S-matrix element of one R-R field strength $F^{(p-2)}$ and one NS-NS vertex operators should not have four-derivative terms.  It has been observed in \cite{Garousi:2010ki} that the disk-level S-matrix element of one R-R and one NS-NS vertex operators produce no such term at order $\alpha'^2$. On the other hand, it has been observed in \cite{Garousi:2006zh} that the low energy expansion of $RP^2$-level and disk-level S-matrix element of two massless closed string  vertex operators are the same  at order $\alpha'^2$, up to an overall factor.   

The disk-level S-matrix element of one R-R potential $C^{(p-3)}$ and two B-field vertex operators has been calculated in \cite{Garousi:2011ut,Becker:2011ar}  from which the couplings of one $F^{(p-2)}$ and two H has been found for D$_p$-brane.  The orintifold projection of the couplings found in \cite{Becker:2011ar} are the same as the above couplings with the following coefficients:
\beqa
b_1=b_7=0,&& b_2=-b_4=b_5=\frac{1}{2}\labell{Sb}
\eeqa
where we have also used the Bianchi identity $dH=0$ to relate the couplings found in \cite{Becker:2011ar} to the couplings in \reef{Lp-2}. We will see that exactly the same coefficients \reef{Sb} are reproduced by the T-duality constraint. This observation and the observation made in \cite{Garousi:2006zh} may indicate that the orientifold projection of the disk-level S-matrix elements at order $\alpha'^2$ are the same as the corresponding $RP^2$-level S-matrix elements at order $\alpha'^2$, up to overall factors.   

The  independent couplings \reef{Lp-2}, however, are not  the most general gauge invariant   couplings because they do not include the Riemann curvature. The gauge invariant couplings involving the Riemann curvature  are the couplings in the CS action \reef{Tf2} which are found by the anomaly cancellation mechanism. The T-duality constraint should reproduce these couplings as well. Hence, we include in this subsection  the following gauge invariant couplings with arbitrary coefficients:
\beqa
{\cal L}_{\rm CS}^{(p-3)}& = &\epsilon^{a_0\cdots a_p}\bigg[\frac{\alpha_1}{(p-3)!}{ C}^{(p-3)}_{a_4\cdots a_{p}}R_{a_{0}a_{1}}{}^{ij}R_{a_{2}a_3\,ij}+\frac{\alpha_2}{(p-3)!}{ C}^{(p-3)}_{a_4\cdots a_{p}}R_{a_{0}a_{1}}{}^{ab}R_{a_{2}a_3\,ab}\bigg]\labell{Tf21}
\eeqa
The two parameters $\alpha_1,\alpha_2$ which are known from the anomaly cancellation mechanism and also from the S-matrix calculation, should be fixed by the T-duality constraint as well.

\subsection{$n=p$ case}

To find all  gauge invariant and  independent couplings involving one  R-R field strength $F^{(p)}$,  we first consider all contractions of one $\epsilon^{a_0\cdots  a_p}$, one $F$, $\nabla F$ or $\nabla\nabla F$, odd number of $H$, $ \nabla H$ and $\nabla\nabla H$, and any number of $\nabla\Phi$, $\nabla\nabla\Phi$, $\nabla\nabla\nabla\Phi$, $R,\,\nabla R$ at four-derivative order. We  remove the terms which are forbidden for O-plane and impose  the  equations of motion. We then impose  the total derivative terms, use the Bianchi identities and $\epsilon$-tensor identities with the same strategy that is discussed in the previous subsection.
In this manner one finds  28 independent couplings.  One particular form for  them is the following\footnote{If one does not use the $\epsilon$-tensor identities, then one would find 46 independent couplings.}: 
\beqa
{\cal L}^{(p)}&\!\!\!\!\!=\!\!\!\!\!\!&\epsilon^{a_{0}...a_{p}}\Big[ 
\frac{c_2}{(p-1)!}\, \nabla_{a}F_{ia_{2}...a_{p}}\, \nabla_{a_{0}}H^{ia}\,_{a_{1}} +\frac{c_3}{(p-1)!}\, \nabla_{a}F_{ia_{2}...a_{p}}\, \nabla^{a}H^{i}\,_{a_{0}a_{1}}\nn\\
&&
  +\frac{c_5}{p!}\, \nabla_{i}F_{a_{1}...a_{p}}\, H^{ia}\,_{a_{0}}\, \nabla_{a} \Phi
 +\frac{c_7}{(p-1)!}\, F_{ia_{2}...a_{p}}\, \nabla^{a}H^{i}\,_{aa_{1}}\, \nabla_{a_{0}} \Phi\nn\\
&&
+\frac{c_8}{(p-1)!}\, F_{ia_{2}...a_{p}}\, \nabla^{a}H^{i}\,_{a_{0}a_{1}}\, \nabla_{a} \Phi 
+\frac{c_{10}}{(p-1)!}\, F_{ia_{2}...a_{p}}\, H^{i}\,_{a_{0}a_{1}}\, \nabla^{a}\nabla_{a} \Phi \nn\\
&&
+\frac{c_{12}}{(p-1)!}\, F_{ia_{2}...a_{p}}\, H^{i}\,_{aa_{1}}\, \nabla^{a}\nabla_{a_{0}} \Phi +\frac{c_{13}}{(p-1)!}\, F_{ja_{2}...a_{p}}\, H_{ia_{0}a_{1}} \nabla^{i}\nabla^{j} \Phi\nn\\
&&
+\frac{c_{14}}{(p-1)!}\, F_{ia_{2}...a_{p}}\, H^{i}\,_{a_{0}a_{1}}\, \nabla^{a} \Phi\, \nabla_{a} \Phi 
+\frac{c_{16}}{(p-1)!}\, F_{ia_{2}...a_{p}}\, H^{i}\,_{aa_{1}} \nabla_{a_{0}} \Phi \nabla^{a} \Phi\nn\\
&& +\frac{c_{17}}{(p-1)!}\, F_{ja_{2}...a_{p}}\, H^{iab}\, H_{iab}\, H^{j}\,_{a_{0}a_{1}}
 +\frac{c_{21}}{(p-3)!}\, F_{jkl a_{4}...a_{p}}\, H^{ikl}\, H_{ia_{0}a_{1}}\, H^{j}\,_{a_{2}a_{3}}\nn\\
&&
  +\frac{c_{23}}{(p-3)!}\, F_{jkl a_{4}...a_{p}}\, H_{ia_{0}a_{1}}\, H^{i}\,_{a_{2}a_{3}}\, H^{jkl}
+\frac{c_{24}}{(p-1)!}\, F_{ka_{2}...a_{p}}\, H_{iaa_{0}}\, H^{ijk}\, H_{j}\,^{a}\,_{a_{1}} \nn\\
&&
+\frac{c_{28}}{(p-1)!}\, F_{ja_{2}...a_{p}}\, H^{iab}\,  H_{ia a_{0}}\, H^{j}\,_{ba_{1}}  
+\frac{c_{30}}{(p-3)!}\, F_{iaba_{4}...a_{p}}\, H^{iab}\, H^{j}\,_{a_{0}a_{1}}\, H_{ja_{2}a_{3}}\nn\\
&& +\frac{c_{31}}{(p-1)!}\, F_{la_{2}...a_{p}}\, H_{ia_{0}a_{1}}\, H^{i}\,_{jk}\, H^{jkl}
+\frac{c_{32}}{(p-1)!}\, F_{ia_{2}...a_{p}}\, H^{i}\,_{a_{0}a_{1}}\, H^{jkl}\, H_{jkl} \nn\\
&&+\frac{c_{33}}{(p-1)!}\, F_{ia_{2}...a_{p}}\, H^{iab}\, H^{j}\,_{aa_{0}} H_{jba_{1}}
+\frac{c_{34}}{(p-3)!}\, F_{ijka_{4}...a_{p}}\, H^{ia}\,_{a_{0}}\, H^{j}\,_{aa_{1}} H^{k}\,_{a_{2}a_{3}} \nn\\
&&+\frac{c_{35}}{(p-1)!}\, F_{ka_{2}...a_{p}}\, H^{ijk}\, R_{ia_{0}ja_{1}}
 +\frac{c_{37}}{(p-1)!}\, F_{ja_{2}...a_{p}}\, H^{i}\,_{aa_{0}} R_{ia_{1}}\,^{ja}\nn\\
&&
+\frac{c_{38}}{(p-2)!}\, F_{jaa_{3}...a_{p}}\, H_{ia_{0}a_{1}} R^{iaj}\,_{a_{2}} +\frac{c_{39}}{(p-1)!}\, F_{ja_{2}...a_{p}}\, H_{iaa_{0}}\, R^{iaj}\,_{a_{1}}\nn\\
&&
+\frac{c_{40}}{(p-3)!}\, F_{ijka_{4}...a_{p}}\, H^{i}\,_{a_{0}a_{1}}\, R^{j}\,_{a_{2}}\,^{k}\,_{a_{3}} +\frac{c_{43}}{(p-1)!}\, F_{ia_{2}...a_{p}}\, H^{i}\,_{a_{0}a_{1}}\, R^{ab}\,_{ab}\nn\\
&&
+\frac{c_{44}}{(p-1)!}\, F_{ia_{2}...a_{p}}\, H^{iab}\, R_{aa_{0}ba_{1}} 
+\frac{c_{46}}{(p-1)!}\, F_{ia_{2}...a_{p}}\, H^{i}\,_{aa_{1}}\, R^{ab}\,_{a_{0}b}\Big]\labell{LP}
\eeqa
Note that in this case also we have set the coefficients of the terms that have world-volume derivative on the R-R field strength, to be zero. However, in the couplings in the first line we use an integration by part to remove one of the two derivatives on $H$ because in imposing T-duality in the next section one needs to dimensionally reduce the couplings. The reduction of $\nabla F\nabla H$ is much easier to perform than the reduction of $F\nabla\nabla H$. In above equation,      $c_2,\cdots, c_{46}$ are 28  arbitrary coefficients that do not depend on $p$. They may be found by the T-duality constraint. 

The coefficients $c_2,c_3$  has been    fixed by the tree-level S-matrix element of one R-R and one NS-NS vertex operators \cite{Garousi:2010ki}, \ie
\beqa
 c_2=2\,,\,\, c_3=-\frac{1}{2}\labell{c2c3}
\eeqa 
In finding this result we write the two-field terms in   \reef{LP} and the couplings found in \cite{Garousi:2010ki} in terms of independent structures, and then force them  to be the same.

\subsection{$n=p+2$ case}

To find all  gauge invariant and independent couplings involving one  R-R field strength $F^{(p+2)}$,  we  consider all contractions of one $\epsilon^{a_0\cdots  a_p}$, one $F$, $\nabla F$ or $\nabla\nabla F$, even number of $H$ and $ \nabla H$, and any number of $\nabla\Phi$, $\nabla\nabla\Phi$, $\nabla\nabla\nabla\Phi$, $R,\,\nabla R$ at four-derivative order. We then  impose  the  equations of motion,  the O-plane conditions,  the total derivative terms,  and use the Bianchi identities and $\epsilon$-tensor identities with the same strategy that is discussed in  the   subsection 2.1.
In this manner one finds  20 independent couplings.  One particular form for  them is the following\footnote{If one does not use the $\epsilon$-tensor identities, then one would find 53 independent couplings.}: 
\beqa
{\cal L}^{(p+2)}&\!\!\!\!\!=\!\!\!\!\!\!&\epsilon^{a_{0}...a_{p}}\Big[ \frac{d_2}{(p+1)!}\, \nabla_{k}F_{la_{0}...a_{p}}\, H^{ijk}\, H_{ij}\,^{l}+\frac{d_3}{(p+1)!}\, \nabla_{i}F_{ja_{0}...a_{p}}\, H^{iac}\, H^{j}\,_{ac}
\nn\\
&& +\frac{d_9}{(p-1)!}\, \nabla_{i}F_{jk la_{2}...a_{p}}\, H^{i}\,_{a_{0}a_{1}}\, H^{jkl}
+\frac{d_{10}}{(p-1)!}\, \nabla_{j}F_{ik la_{2}...a_{p}}\, H^{i}\,_{a_{0}a_{1}}\, H^{jkl}\nn\\
&&+\frac{d_{11}}{(p+1)!}\, \nabla_{i}F_{ja_{0}...a_{p}}\, \nabla^{i} \nabla^{j} \Phi
+\frac{d_{12}}{(p+1)!}\, \nabla_{i}F_{ja_{0}...a_{p}}\, R^{ia j}\,_{a}\nn\\
&& 
 +\frac{d_{15}}{p!}\,\nabla_{a} F_{ij a_{1}...a_{p}}\,  R^{iaj}\,_{a_{0}}
+\frac{d_{16}}{p!}\, F_{ij a_{1}...a_{p}}\, R^{iaj}\,_{a_{0}}\, \nabla_{a} \Phi\nn\\
&& +\frac{d_{21}}{p!}\, F_{jk a_{1}...a_{p}}\, \nabla^{a}H^{i jk}\, H_{ia a_{0}}
+\frac{d_{22}}{p!}\, F_{jk a_{1}...a_{p}}\, \nabla^{i}H^{jk a}\, H_{ia a_{0}}\nn\\
&& 
+\frac{d_{26}}{p!}\, F_{jk a_{1}...a_{p}}\, \nabla^{a}H_{ia a_{0}}\, H^{ijk} +\frac{d_{27}}{p!}\, F_{kl a_{1}...a_{p}}\, \nabla_{a_{0}}H^{ij k}\, H_{ij}\,^{l}
 \nn\\
&&+\frac{d_{29}}{p!}\, F_{ij a_{1}...a_{p}}\, \nabla_{a}H^{i}\,_{ba_{0}}\, H^{jab}
+\frac{d_{30}}{p!}\, F_{ij a_{1}...a_{p}}\, \nabla_{a_{0}}H^{i}\,_{ab}\, H^{jab}\nn\\
&& 
+\frac{d_{36}}{p!}\, F_{ij a_{1}...a_{p}}\, \nabla_{a}H^{iab}\, H^{j}\,_{ba_{0}} +\frac{d_{41}}{(p-2)!}\, F_{ij kl a_{3}...a_{p}}\, \nabla_{a_{0}}H^{i}\,_{a_{1}a_{2}}\, H^{jk l}
\nn\\
&&+\frac{d_{42}}{p!}\, F_{il a_{1}...a_{p}}\, \nabla^{i}H_{jka_{0}}\, H^{jk l} +\frac{d_{43}}{(p-2)!}\, F_{ij kl a_{3}...a_{p}}\, H^{i}\,_{a_{1}a_{2}}\, H^{jkl}\, \nabla_{a_{0}}  \Phi
 \nn\\
&&+\frac{d_{47}}{p!}\, F_{ij a_{1}...a_{p}}\, H^{iab}\, H^{j}\,_{ba_{0}}\, \nabla_{a} \Phi
+\frac{d_{48}}{p!}\, F_{jk a_{1}...a_{p}}\, H_{ia a_{0}}\, H^{ijk}\, \nabla^{a} \Phi  \Big]\labell{Lp+2}
\eeqa
where the   $p$-independent coefficients $d_2,\cdots, d_{48}$   may be found by the T-duality constraint. 

The coefficients $d_{11},d_{12},  d_{15}$  have been   fixed by the tree-level S-matrix element of one R-R and one NS-NS vertex operators \cite{Garousi:2010ki}. They are
\beqa
d_{11}=-2\,,\,\,d_{12}=-2\,,\,\, d_{15}=2\labell{d1112}
\eeqa
In finding the above result, we  have imposed  the first Bianchi identity in \reef{Bian} on the two-field couplings found in \cite{Garousi:2010ki}. Note that as observed in \cite{Garousi:2010ki} the above results indicate that the curvature $R^{iaj}{}_a$ and $\nabla^i\nabla^j\Phi$ appear in the O-plane action as $ij$-component of the following combination:
\beqa
{\cal R}^{AB}&=&R^{A aB}{}_a+\nabla^A\nabla^B\Phi\labell{cR}
\eeqa
where $A,B$ are 10-dimensional bulk indices. Note that the transverse contraction of the Riemann curvature, \ie $R^{AiB}{}_{i}$ has been removed at the onset by imposing the equations of motion. This dilaton-Riemann curvature  appears also in NS-NS couplings of O-plane action at order $\alpha'^2$ \cite{Garousi:2014oya}. We speculate that the second derivative of dilaton   appears in all O-plane and D-brane couplings  in above combination.

\subsection{$n=p+4$ case}

Performing the same steps as in subsection 2.1, one finds   there are 19  independent couplings on the world-volume of O$_p$-plane  that are not related to each other by the  Bianchi identities, $\epsilon$-tensor identities  and the  total derivative terms.  One particular form for  the  couplings is the following\footnote{If one does not use the $\epsilon$-tensor identities, then one would find 47 independent couplings.}: 
\beqa
{\cal L}^{(p+4)}&\!\!\!\!\!=\!\!\!\!\!\!&\epsilon^{a_{0}...a_{p}}\Big[\frac{e_1}{(p+1)!}\nabla_{a} F_{ijk a_{0}...a_{p}}\,\nabla^{a}H^{ijk}
 +\frac{e_3}{(p+1)!}\, F_{ijka_{0}...a_{p}}\,  \nabla^{a}H^{ijk}\, \nabla_{a}\Phi\nn\\
&&
 +\frac{e_6}{(p+1)!}\, F_{ijk a_{0}...a_{p}}\, H^{ijk}\, \nabla^{a}\nabla_{a}\Phi 
+\frac{e_8}{(p+1)!}\, F_{jkla_{0}...a_{p}}\, H^{ikl}\, \nabla^{j}\nabla_{i}\Phi \nn\\
&&+\frac{e_9}{(p+1)!}\, F_{ijk a_{0}...a_{p}}\, H^{ijk}\, \nabla^{a}\Phi\, \nabla_{a}\Phi
 +\frac{e_{12}}{(p+1)!}\,  F_{kmna_{0}...a_{p}} H^{ijk} H_{i}\,^{lm} H_{jl}\,^{n} \nn\\
&&+\frac{e_{13}}{(p-1)!}\,  F_{iklmna_{2}...a_{p}}\, H^{i}\,_{a_{0}a_{1}}\, H^{j kl}\, H_{j}\,^{mn}
 +\frac{e_{17}}{(p+1)!}\,  F_{jkla_{0}...a_{p}}\, H^{iab}\, H_{iab}\, H^{jkl}\nn\\
&&
 +\frac{e_{20}}{(p+1)!}\,  F_{jkla_{0}...a_{p}}\, H_{iab}\, H^{ikl}\, H^{jab} 
+\frac{e_{26}}{(p-1)!}\,  F_{ijklma_{2}...a_{p}}\, H^{ia}\,_{a_{0}}\, H^{j}\,_{aa_{1}}\, H^{klm} \nn\\
&&
+\frac{e_{28}}{(p+1)!}\, F_{ijka_{0}...a_{p}}\, H^{iab}\, H^{j}\,_{ac}\, H^{k}\,_{b}\,^{c} +\frac{e_{31}}{(p+1)!}\,  F_{lmna_{0}...a_{p}}\, H_{ijk}\, H^{ijl}\, H^{kmn}\nn\\
&&
+\frac{e_{32}}{(p-1)!}\, F_{jklmna_{2}...a_{p}}\, H_{ia_{0}a_{1}}\, H^{ijk}\, H^{lmn} +\frac{e_{33}}{(p+1)!}\, F_{lmna_{0}...a_{p}}\, H^{ijk}\, H_{ijk}\, H^{lmn}\nn\\
&&
 +\frac{e_{35}}{(p+1)!}\,  F_{klma_{0}...a_{p}}\, H_{ij}\,^{k}\, R^{il jm}
 +\frac{e_{37}}{(p+1)!}\,  F_{jkla_{0}...a_{p}}\, H^{ijk}\, R^{la}\,_{ia} \nn\\
&& 
+\frac{e_{42}}{(p+1)!}\,  F_{ijka_{0}...a_{p}}\, H^{i}\,_{ab}\, R^{ja kb}
+\frac{e_{44}}{(p-1)!}\,  F_{ijklma_{2}...a_{p}}\, H^{ijk}\, R^{l}\,_{a_{0}}\,^{m}\,_{a_{1}} \nn\\
&& +\frac{e_{47}}{(p+1)!}\,  F_{ijka_{0}...a_{p}}\, H^{ijk}\, R^{ab}\,_{ab}\Big]\labell{Lp+4}
\eeqa
where   the   $p$-independent coefficients $e_1,\cdots, e_{47}$   may be found by the T-duality constraint. 

The coefficient $e_{1}$  has been   fixed by the tree-level S-matrix element of one R-R and one NS-NS vertex operators \cite{Garousi:2010ki}, \ie  
\beqa
e_1=-\frac{1}{3!}\labell{ee}
\eeqa
The proposal  that the combination \reef{cR} should appear in the world-volume couplings, dictates that   the T-duality should fix  the coefficient $e_8$  to be the same as $e_{37}$. As we will see in section 4, the T-duality indeed produces this relation.

\subsection{$n=p+6$ case}

Similar calculation for the couplings involving one R-R field strength $F^{(p+6)}$ gives the following two independent coupling:
\beqa
 {\cal L }^{(p+6)}&\!\!\!=\!\!\!\!& \epsilon^{a_{0}...a_{p}}\Big[\frac{f_1}{p!}\, F_{ij kl mn a_{1}...a_{p}}\,\nabla_{a_{0}}H^{i jk}\, H^{lm n}
 +\frac{f_2}{(p+1)!}\, \nabla_{i}F_{jk lmn a_{0}...a_{p}}\, H^{ijk}\, H^{lmn}\Big]\labell{LP+6}
\eeqa
where   $f_1,f_{2}$ are two arbitrary coefficients that may be found by the T-duality constraint.  There are no couplings involving one NS-NS field which is consistent with   the tree-level S-matrix element of one R-R and one NS-NS vertex operators \cite{Garousi:2010ki}. The above two coefficients may be fixed by the low energy expansion of $RP^2$-plane S-matrix element of one R-R and two NS-NS vertex operators at order $\alpha'^2$. The disk-level calculations have been fixed these coefficients to be zero  \cite{Garousi:2011ut}. We will see that the T-duality also fix these coefficients for O-plane to be zero which is consistent with the speculation that the orientifold projection of D-brane couplings at order $\alpha'^2$ is the same as O-plane couplings at order $\alpha'^2$, up to overall factors.

Therefore,   there are 76 independent couplings at order $\alpha'^2$ which have one R-R field. These gauge invariant couplings  are the appropriate couplings on the world-volume of O$_p$-plane for some specific values for the 76 parameters. They may be found by the S-matrix or other methods in string theory. We are going to find these parameters in this paper  by the T-duality constraint. We will find that  all 76 parameters are fixed up to an overall factor. 

\section{T-duality transformations}

When compactifying the superstring theory on a circle with radius $\rho$ and  with the coordinate $y$, the full nonlinear  T-duality transformations at the leading order of $\alpha'$ for the NS-NS and R-R  fields   are given in  \cite{Buscher:1987sk,Buscher:1987qj,Meessen:1998qm}, \ie
\beqa
e^{2\phi'}=\frac{e^{2\phi}}{G_{yy}}&;& 
G'_{yy}=\frac{1}{G_{yy}}\nonumber\\
G'_{\mu y}=\frac{B_{\mu y}}{G_{yy}}&;&
G'_{\mu\nu}=G_{\mu\nu}-\frac{G_{\mu y}G_{\nu y}-B_{\mu y}B_{\nu y}}{G_{yy}}\nonumber\\
B'_{\mu y}=\frac{G_{\mu y}}{G_{yy}}&;&
B'_{\mu\nu}=B_{\mu\nu}-\frac{B_{\mu y}G_{\nu y}-G_{\mu y}B_{\nu y}}{G_{yy}}\labell{nonlinear}\\
C'^{(n)}_{\mu\cdots \nu \alpha y}&=& C^{(n-1)}_{\mu\cdots \nu \alpha}- \frac{C^{(n-1)}_{[\mu\cdots\nu|y}G_{|\alpha]y}}{G_{yy}}\,\,;\nn\\
C'^{(n)}_{\mu\cdots\nu\alpha\beta}&=&C^{(n+1)}_{\mu\cdots\nu\alpha\beta y}+ C^{(n-1)}_{[\mu\cdots\nu\alpha}B_{\beta]y}+ \frac{C^{(n-1)}_{[\mu\cdots\nu|y}B_{|\alpha|y}G_{|\beta]y}}{G_{yy}}\nn
\eeqa
 where $\mu,\nu$ denote any   direction other than $y$.  Our notation for making  antisymmetry  is such that \eg $C^{(2)}_{[\mu_1\mu_2}B_{\mu_3]\nu}=C^{(2)}_{\mu_1\mu_2}B_{\mu_3\nu}-C^{(2)}_{\mu_3\mu_2}B_{\mu_1\nu}+C^{(2)}_{\mu_3\mu_1}B_{\mu_2\nu}$. In above transformations the metric is  in the string frame.  If one assumes fields are transformed covariantly under the coordinate transformations, then the above transformations receive  corrections at order $\alpha'^3$ in the superstring theory \cite{Razaghian:2018svg} in which we are not interested because the couplings in this paper are   at order $\alpha'^2$.

To impose the T-duality constraint on the effective action, one should first  write  all independent gauge invariant couplings of O$_p$-plane, as we have done in the previous section,  and then reduce them on the circle when O$_p$-plane is along the circle. The  T-duality transformation of the reduced action should be the same as the reduction of O$_{p-1}$-plane when it is orthogonal to the circle,   up to some total derivative terms.     To impose the T-duality constraint on the effective  action, however, it is  convenient to use the following reductions for  the metric, $B$-field, dilaton and the R-R potentials \cite{Maharana:1992my,Garousi:2019jbq}:
  \beqa
G_{AB}&=&\left(\matrix{\bg_{\mu\nu}+e^{\varphi}g_{\mu }g_{\nu }& e^{\varphi}g_{\mu }&\cr e^{\varphi}g_{\nu }&e^{\varphi}&}\right),\, B_{AB}= \left(\matrix{\bb_{\mu\nu}+\frac{1}{2}b_{\mu }g_{\nu }- \frac{1}{2}b_{\nu }g_{\mu }&b_{\mu }\cr - b_{\nu }&0&}\right)\nn\\ \Phi&=&\bar{\phi}+\varphi/4\labell{reduc}\\
C^{(n)}_{\mu_1\cdots\mu_n}&=&\bc^{(n)}_{\mu_1\cdots\mu_n}+\bc^{(n-1)}_{[\mu_1\cdots\mu_{n-1}}g_{\mu_n]}\nn\\
C^{(n)}_{\mu_1\cdots \mu_{n-1}y}&=&\bc^{(n-1)}_{\mu_1\cdots\mu_{n-1}}\nn
\eeqa
where $\bg_{\mu\nu}, \bb_{\mu\nu}, \bar{\phi} $ and $\bc^{(n)}$  are the metric,      B-field,  dilaton and the  R-R potentials, respectively,  in the $9$-dimensional base space, and $g_{\mu},\, b_{\mu}$ are two vectors  in this space.  
In this parametrization, inverse of metric becomes
\beqa
G^{AB}=\left(\matrix{\bg^{\mu\nu} &  -g^{\mu }&\cr -g^{\nu }&e^{-\varphi}+g_{\alpha}g^{\alpha}&}\right)\labell{inver}
\eeqa
where $\bg^{\mu\nu}$ is the inverse of the base  metric which raises the indices of  the    vectors.
The nonlinear T-duality transformations \reef{nonlinear}   in the parametrizations  \reef{reduc} then  become  remarkably the following linear transformations:
\beqa
\varphi'= -\varphi
\,\,\,,\,\,g'_{\mu }= b_{\mu }\,\,\,,\,\, b'_{\mu }= g_{\mu } \labell{T2}
\eeqa
and all other $9$-dimensional    fields remain invariant under the T-duality transformation. Note that the T-duality transformation of the base space R-R potential $\bc^{(n)}$ is trivial in the parametrization \reef{reduc}, however,  the R-R gauge transformation of this potential   in which we are not interested in this paper, seems to be non-trivial.

One can easily verify that the CS action at order $\alpha'^0$  is invariant under the T-duality. If the killing coordinate $y$ is a world volume, then  the T-duality transformation of the reduction of O$_p$-plane action in the parametrization \reef{reduc} becomes
\beqa
 T_{p-1}\int d^p x\,\epsilon^{a_0\cdots a_{p-1}}\frac{1}{p!}\bc^{(p)}_{a_0\cdots a_{p-1}}\labell{TOp}
\eeqa
where we have used the relation $2\pi \rho T_p=T_{p-1}$ and $\epsilon^{a_0\cdots a_{p-1}y}=\epsilon^{a_0\cdots a_{p-1}}$. On the other hand, the reduction of the  O$_{p-1}$-plane action  in the parametrization \reef{reduc} when the $y$-coordinate is  transverse to the  O$_{p-1}$-plane is 
\beqa
 T_{p-1}\int d^p x\,\epsilon^{a_0\cdots a_{p-1}}\frac{1}{p!}\bigg(\bc^{(p)}_{a_0\cdots a_{p-1}}+p\bc^{(p-1)}_{[a_0\cdots a_{p-2}}g_{a_{p-1}]}\bigg)
\eeqa
Using the fact that $g_{a_{p-1}}$ is the component of the 10-dimensional metric which has one $y$-index and $y$ is a transverse index in this case, the last term above is removed for the  O-plane. The rest is the same as the action \reef{TOp}.

There is no such symmetry for the CS action at higher orders of $\alpha'$ because the Riemann curvature is not invariant under the T-duality transformations.  As a result, one has to add some other terms to this action to make it T-duality invariant as in the leading order term. Since the new couplings involve R-R and NS-NS field strengths and their covariant derivatives, it is convenient to first  find the reduction of these field strengths and then apply them to find the reduction of each gauge invariant coupling. 

 Using the reductions \reef{reduc}, it is straightforward to   calculation reduction of the Riemann curvature, $H$, $\nabla H$, $\nabla\Phi$ or $\nabla\nabla\Phi$. As it has been argued in \cite{Garousi:2019mca}, after writing the reductions in terms of $\bH$ which is defined as
 \beqa
  \bH=d\bb-\frac{1}{2}g\wedge W-\frac{1}{2}b\wedge V
 \eeqa
where $W=db$ and $V=dg$, they have  two parts. One part includes terms which are invariant under $U(1)\times U(1)$ gauge transformations corresponding to the gauge fields $g_\mu,\,b_{\mu}$. They have  been found in \cite{Garousi:2019mca} (see eq.(35), eq.(36) and eq.(37) in this reference\footnote{There is a typo in the reduction of $\nabla_\mu H_{\nu\alpha y}$ in eq.(37) in the published version of \cite{Garousi:2019mca}. The first term on the right hand side of this expression should be negative.}). The other part which is not invariant under the  $U(1)\times U(1)$ gauge transformations,  includes  the gauge fields $g_\mu,\,b_{\mu}$ without derivative on them. Such terms are cancelled at the end of the day in the reduction of a 10-dimensional gauge invariant coupling. So one may  keep only  the $U(1)\times U(1)$ gauge invariant parts of the reduction of the Riemann curvature, $H$, $\nabla H$, $\nabla\Phi$ and $\nabla\nabla\Phi$, and the following   reduction of the inverse of the  spacetime metric:
 \beqa
G^{AB}&=&\left(\matrix{\bg^{\mu\nu} &0&\cr 0&e^{-\varphi}&}\right)\labell{GAB}
\eeqa
and removes all other terms in the reduction. In this way one can find the reduction of any gauge invariant bulk coupling. However, the   metric $G^{AB}$ is not used in constructing the O$_p$-plane couplings in the previous section.  The   world-volume couplings in fact   are constructed by contracting the  tensors       with the first fundamental form $\tG^{AB}=\prt_a X^A\prt_b X^B\tg^{ab}$ which projects the spacetime tensors to the world-volume directions, and with  $\bot^{AB}=G^{AB}-\tG^{AB}$ which projects the tensor to the transverse directions.  In the first fundamental form, $\tg^{ab}$ is inverse of the pull-back metric $\tg_{ab}=\prt_a X^A\prt_b X^B G_{AB}$.

In the static gauge where $X^{a}=\sigma^a$ and for the O$_p$-plane at $X^i=0$, one has $\tG^{ij}=\tG^{ai}=\tG^{ia}=0$, and $\tG^{ab}=\tg^{ab}$, $\tg_{ab}=G_{ab}$.  When O$_p$-plane is orthogonal to the   killing coordinate, the  first fundamental form and world-volume components of the inverse  of the spacetime metric  have no component along the $y$-direction,  because $y$ is a transverse direction. Hence, in this case $\bot^{ab}=0$. Moreover $\bot^{ai}=G^{ai}=0$ by  the orientifold projection. The non-zero components in this case are 
\beqa
\tG^{ab}=G^{ab}=\bg^{ab},&& \bot^{ij}=G^{ij}= \left(\matrix{\bg^{\ti\tj} &0\cr0&e^{-\varphi}&}\right)\labell{Gab}
\eeqa
The gauge field  $g_{\ta}$ does not appear in  $\tG^{ab}$, however, it  appears in the reduction of $\bot^{ij}$. As in \reef{GAB}, we have ignored it because we have ignored the non-gauge invariant  terms in the reduction of the Riemann curvature, $H$, $\nabla H$, $\nabla\Phi$ and $\nabla\nabla\Phi$.

On the other hand, when O$_p$-plane is along the killing coordinate,  both the  first fundamental form and world-volume components of the inverse  of the spacetime metric  have   component along the $y$-direction, however, because $\tG^{ab}=G^{ab}$ one again has $\bot^{ab}=0$.    In this case   the non-zero components   are 
\beqa
\tG^{ab}=G^{ab}=\left(\matrix{\bg^{\ta\tb} &0&\cr 0&e^{-\varphi}&}\right),&& \bot^{ij}=G^{ij} =\bg^{ij} \labell{Gabw}
\eeqa
The gauge field  $g_{\ta}$ does not appear in $\bot^{ij}$, however, it appears in the reduction of $\tG^{ab}$ that we have again removed it.

Using the  reduction of the R-R potential in \reef{reduc}, one can find the reduction of R-R field strength and its first derivative which appear in the couplings in the previous section. They have again   two parts. One part is not invariant under the  $U(1)\times U(1)$ gauge transformations which is cancelled in the gauge invariant couplings, hence we ignore it. The   $U(1)\times U(1)$ gauge invariant part of the  reduction is  
\beqa
F^{(n)}_{\mu_{1}...\mu_{n-1}y}&=& \frac{}{}\bar{F}^{(n-1)}_{\mu_{1}...\mu_{n-1}}
+(-1)^{(n-3)}\,\, W_{[\mu_{1}\mu_{2}}\,\, \bar{c}^{(n-3)}_{\mu_{3}...\mu_{n-1}]}
+\bar{H}_{[\mu_{1}\mu_{2}\mu_{3}} \bar{c}^{(n-4)}_{\mu_{4}...\mu_{n-1}]}\equiv  {F^W}^{(n-1)}_{\mu_{1}...\mu_{n-1}}\nn\\
F^{(n)}_{\mu_{1}...\mu_{n}}&=& \frac{}{}\bar{F}^{(n)}_{\mu_{1}...\mu_{n}}
+(-1)^{(n-2)}\, V_{[\mu_{1}\mu_{2}}\,\, \bar{c}^{(n-2)}_{\mu_{3}...\mu_{n}]}
+\bar{H}_{[\mu_{1}\mu_{2}\mu_{3}}\,\, \bar{c}^{(n-3)}_{\mu_{4}...\mu_{n}]}\equiv  {F^V}^{(n)}_{\mu_{1}...\mu_{n}}\nn\\
 \nabla_{y}F^{(n)}_{\mu_{1}...\mu_{n-1}y}&\!\!\!\!\!=\!\!\!\!\!&\frac{1}{2} e^{\varphi}\Bigg[ {F^V}^{(n)}_{\mu_{1}...\mu_{n-1}\mu}\,\, \nabla^{\mu}\varphi
-{F^W}^{(n-1)}_{[\mu\mu_{2}...\mu_{n-1}} V_{\mu_{1}]}{}^{\mu} \Bigg]\nn\\	
\nabla_{y}F^{(n)}_{\mu_{1}...\mu_{n}}&\!\!\!\!\!=\!\!\!\!\!&-\frac{1}{2}\Bigg[ (-1)^{(n-1)} {F^W}^{(n-1)}_{[\mu_{2}...\mu_{n}}\nabla_{\mu_{1}]}\varphi+e^{\varphi}\,  {F^V}^{(n)}_{[\mu\mu_{2}...\mu_{n}}\, V_{\mu_{1}]}{}^{\mu}
 \Bigg]\nn\\
\nabla_{\nu}F^{(n)}_{\mu_{1}...\mu_{n}}&\!\!\!\!\!=\!\!\!\!\!&\frac{1}{2}\Bigg[2\nabla_{\nu} {F^V}^{(n)}_{\mu_{1}...\mu_{n}}
  - (-1)^{(n-1)} {F^W}^{(n-1)}_{[\mu_{2}...\mu_{n}} V_{\mu_{1}]\nu}\Bigg]\nn\\
\nabla_{\nu}F^{(n)}_{\mu_{1}...\mu_{n-1}y}&\!\!\!\!\!=\!\!\!\!\!&\frac{1}{2}\Bigg[2\nabla_{\nu} {F^W}^{(n-1)}_{\mu_{1}...\mu_{n-1}}
  +  e^{\varphi} {F^V}^{(n)}_{\mu_{1}...\mu_{n-1}\mu}\, V^{\mu}{}_{\nu}-  {F^W}^{(n-1)}_{\mu_{1}...\mu_{n-1}}\,\, \nabla_{\nu}\varphi 
 \Bigg]
\eeqa
where the covariant derivatives on the right-hand side are 9-dimensional and $\bar{F}=d\bc$. One can check that the reduction of $\nabla H$ found in \cite{Garousi:2019mca} can be found from the  above reduction   when one uses $H^{W(2)}=W$ and $H^{V(3)}=\bH$. Obviously, the 
   $U(1)\times U(1)$ gauge invariant part of the reduction of   the R-R potential $C$ is 
\beqa
C^{(n)}_{\mu_1\cdots\mu_n}&=&\bc^{(n)}_{\mu_1\cdots\mu_n} \nn\\
C^{(n)}_{\mu_1\cdots \mu_{n-1}y}&=&\bc^{(n-1)}_{\mu_1\cdots\mu_{n-1}}\labell{FC}
\eeqa
Using the above  $U(1)\times U(1)$ gauge invariant part of the reductions, one can calculate the reduction  of any 10-dimensional gauge invariant coupling. The result would be the same as writing the coupling in terms of ordinary derivatives of metric, B-field, dilaton and R-R potential  and then using the reductions \reef{reduc}. For example, using the above reduction for the R-R field strength, one finds the following reduction for the gauge invariant coupling $F^2$:
\beqa
\frac{1}{n!}F^{(n)}\cdot F^{(n)}&=&\frac{1}{n!}F^{V(n)}\cdot F^{V(n)}+\frac{e^{-\varphi}}{(n-1)!}F^{W(n-1)}\cdot F^{W(n-1)}
\eeqa
which is the correct reduction that has been found in \cite{Garousi:2019jbq}  by writing the R-R field strength in terms of R-R potential and using the reductions \reef{reduc}. It is obvious that the left-hand side is invariant under the 10-dimensional R-R gauge transformations, hence, the right-hand side should be also invariant under the 9-dimensional R-R gauge transformations. This might be used to define the gauge transformation of the base space R-R potential $\bc^{(n)}$ in which we are not interested in this paper.
 
 As another  example,  the O$_p$-plane world-volume reduction of the CS terms  in \reef{Tf2} are
\beqa
&&\epsilon^{a_{0}...a_{p}}\frac{1}{(p-3)!} C^{(p-3)}_{a_{4}...a_{p}}\, R_{a_{0}a_{1}ij}\, R_{a_{2}a_{3}}\,^{ij}=\nn\\
&& \epsilon^{a_{0}...a_{p-1}}\, e^{2\varphi}\Bigg[\frac{1}{4(p-4)!}\, \bar{c}^{(p-4)}_{a_{4}...a_{p-1}}\, V_{a_{0}a_{1}}\, V_{a_{2}a_{3}}\, V_{ij}\, V^{ij}\nn\\
&&\qquad\qquad\qquad\qquad \qquad\qquad\qquad\quad-\frac{1}{(p-3)!}\bar{c}^{(p-3)}_{a_{3}...a_{p-1}}\Big( \nabla_{a_{0}}\varphi V_{a_{1}a_{2}} V_{ij} V^{ij}
+ \nabla_{a_{0}}V_{ij}V_{a_{1}a_{2}} V^{ij}\Big)\Bigg]\nn\\
&&\epsilon^{a_{0}...a_{p}}\frac{1}{(p-3)!} C^{(p-3)}_{a_{4}...a_{p}}\, R_{a_{0}a_{1}ab}\, R_{a_{2}a_{3}}\,^{ab}=\nn\\
&& \epsilon^{a_{0}...a_{p-1}} e^{\varphi}\Bigg[\frac{1}{(p-3)!}\bar{c}^{(p-3)}_{a_{3}...a_{p-1}}\Big(
e^{\varphi} \nabla_{a_{0}} V_{ab} V^{a}_{a_{1}}  V^{b}_{a_{2}} -e^{\varphi} \nabla_{a_{0}} V_{cd} V^{cd} V_{a_{1}a_{2}}
-e^{\varphi} \nabla_{a} V_{a_{0}a_{1}} V^{a}_{b} V^{b}_{a_{2}}\nn\\
&&
\qquad\qquad\quad-e^{\varphi}\, V_{cd}\, V^{cd}\, V_{a_{1}a_{2}}\, \nabla_{a_{0}} \varphi-2\, e^{\varphi}\,  V_{ab}\, V^{a}\,_{a_{1}}\, V^{b}\,_{a_{2}}\, \nabla_{a_{0}} \varphi
- \nabla^{a} V_{a_{1}a_{2}}\, \nabla_{a} \varphi\, \nabla_{a_{0}} \varphi\nn\\&&\qquad\qquad\quad-2\, V_{a_{1}a_{2}}\, \nabla_{a}\nabla_{a_{0}}\varphi
+2\, V^{a}\,_{a_{2}}\,  \nabla_{a}\nabla_{a_{1}}\varphi\, \nabla_{a_{0}}\varphi -2\, V_{a_{1}a_{2}}\,  \nabla_{a}\nabla_{a_{0}}\varphi\, \nabla^{a} \varphi\nn\\&&
\qquad\qquad\quad- V_{a_{1}a_{2}}\, \nabla_{a} \varphi\, \nabla^{a} \varphi\, \nabla_{a_{0}} \varphi\Big) +\frac{1}{(p-4)!}\bar{c}^{(p-4)}_{a_{4}...a_{p-1}}\Big(
\frac{1}{4}\, e^{\varphi}\,  V_{ab}\, V^{ab}\, V_{a_{0}a_{1}}\, V_{a_{2}a_{3}}\nn\\&&
\qquad\qquad\quad
+\frac{1}{2}\, e^{\varphi}\, V_{ab}\, V^{a}\,_{a_{0}}\, V^{b}\,_{a_{1}}\, V_{a_{2}a_{3}} +\frac{1}{2}\, \nabla_{a} V_{a_{2}a_{3}}\, \nabla^{a} V_{a_{0}a_{1}}
+ \nabla_{a} V_{a_{1}a_{2}}\, V^{a}\,_{a_{3}} \nabla_{a_{0}} \varphi\nn\\&&
\qquad\qquad\quad
+\nabla^{a} V_{a_{0}a_{1}}\, V_{a_{2}a_{3}}\, \nabla \varphi_{a} +V^{a}\,_{a_{1}}\, V_{a_{2}a_{3}}\, \nabla_{a} \varphi\, \nabla_{a_{0}} \varphi
+\frac{1}{2}\, V_{a_{0}a_{1}}\, V_{a_{2}a_{3}}\, \nabla_{a} \varphi\, \nabla^{a} \varphi\Big)\Bigg]\labell{RRW}
\eeqa
In finding the above result we have separated  the world-volume indices to $y$ and the world indices which do not include the $y$-index, then we have used  the reduction for each tensors. We have   assumed the 9-dimensional base space is flat, and removed the terms that are projected out by the orientifold projection, \eg we have removed $V_{ai}$ because $g_i$ is related to   $G_{iy}$ and $y$ is world-volume index, hence, it is projected out. Note that the world-volume indices on the right-hand side do not include the $y$-index.

The O$_{p-1}$-plane  transverse reduction of the CS  terms   are
\beqa
&&\epsilon^{a_{0}...a_{p-1}}\frac{1}{(p-4)!} C^{(p-4)}_{a_{4}...a_{p-1}}\, R_{a_{0}a_{1}ij}\, R_{a_{2}a_{3}}\,^{ij} =
\nn\\ &&\qquad\qquad\qquad\qquad \qquad \epsilon^{a_{0}...a_{p-1}}\frac{ e^{\varphi}}{(p-4)!}\bar{c}^{(p-4)}_{a_{4}...a_{p-1}}\Bigg[\frac{1}{2} \nabla_{i}V_{a_{0}a_{1}} \nabla^{i}V_{a_{2}a_{3}}
-\nabla_{i} V_{a_{1}a_{2}}\, V_{a_{3}}\,^{i}\, \nabla_{a_{0}}\varphi\Bigg]\nn\\
&&\epsilon^{a_{0}...a_{p-1}}\frac{1}{(p-4)!} C^{(p-4)}_{a_{4}...a_{p-1}}\, R_{a_{0}a_{1}ab}\, R_{a_{2}a_{3}}\,^{ab}=0\labell{RRT}
\eeqa
In finding the above result we have separated  the transverse  indices to $y$ and the transverse indices which do not include the $y$-index, then we have used  the reduction for each tensors. Here,  we have also removed the terms that are projected out for O-plane, \eg we have removed $V_{ab}$ because $g_a$ is related to   $G_{ay}$ and $y$ is transverse index, hence, it is projected out. Note that the transverse indices on the right-hand side do not include the $y$-index. Similar calculations as above can be done for all couplings in the previous section. Writing the reduced couplings in terms of the base fields $\bc,\, V,\cdots$, one can easily transform them under the T-duality transformations \reef{T2}.

\section{T-duality constraint on the   couplings}

It has been observed in  \cite{Garousi:2019wgz,Garousi:2019mca} that the T-duality constraints on the couplings in the bosonic string theory at order $\alpha'$ and $\alpha'^2$ are the same whether  or not   the base space is flat. In fact, the constraints that one finds between the coefficients of effective action when base space is flat are exactly the same  constraints as one finds for the curved base space. So it is convenient to    consider the reduction  of the couplings in   section 2 on the flat base space, and then impose the T-duality constraint on them to find the unknown coefficients of the couplings. 

The T-duality constraint is   
\beqa
{ \Delta}-{\cal J}&=&0\labell{OpOp}
\eeqa 
where $\Delta=$O$_{(p-1)}$-plane-(O$_{p}$-plane)$'$. The first term in $\Delta$ is transverse reduction of O$_{(p-1)}$-plane and the second term is T-duality of world-volume reduction of O$_p$-plane. The ${\cal J}$ in above equation  represents some  total derivative terms in the flat base space, \ie
\beqa
{\cal J}^n&=&\int d^p\, \bg^{ab}\prt_a{\cal I}_b^n
\eeqa
where the vector ${\cal I}_a^n$ is made of $\epsilon^{a_0\cdots a_{p-1}}$ and the base space fields, $\bc^{(n)}$, $V,W,\bH$, $\prt\varphi, \prt\bar{\phi}$ and their derivatives at three derivative orders. Moreover, to produce the the same structures that appear in $\Delta$, one should multiply each $WW$ or its derivatives by factor $e^{-\varphi}$, each $VV$  by factor $e^{\varphi}$, each extra $W$  by factor $e^{-\varphi}$ and each extra $V$ or $VW$  with no such factor. These factors are traced to the parametrisation  we have used in the reductions \reef{reduc}. 

The T-duality constraint \reef{OpOp} is similar to the equation \reef{DL}. Hence, to solve it one should use the following Bianchi identities for the field strengths $V,W, \bH$:
\beqa
dW=0\,\,;\,\, dV=0\,\,;\,\, d\bH=-\frac{3}{2}W\wedge V\labell{WVH}
\eeqa
and should use the $\epsilon$-tensor identities. Here also we find that it is easy to impose the above Bianchi identities by writing the field strengths $W,V$ or $\bH$, in terms of the potentials $b_{\mu},g_{\mu}, \bb_{\mu\nu}$. Moreover, to impose the $\epsilon$-tensor identities, we write  the resulting non-gauge invariant couplings   explicitly in terms of the   values that each world-volume  index can take, \eg $a_0=0,1,2,\cdots,p-1$. Performing these steps, one rewrites the equation \reef{OpOp} in terms of independent structures. Solving them then one finds  the parameters of the gauge invariant  couplings found in section 2. This is the strategy that we follow in this section.


 To impose the constraint \reef{OpOp}, we  note   that the reduction of $F^{(n)}$, involves the base space fields $\bc^{(n-1)},\, \bc^{(n-2)},\, \bc^{(n-3)}$ and $\bc^{(n-4)}$. So the world-volume reduction of O$_p$-plane  and the transverse reduction of O$_{p-1}$-plane produces the following 9-dimensional R-R potentials: 
\beqa
  F^{(p+6)}\,\rightarrow &&\,\left\{\matrix{O_p\,\,\,\,\,:&\bc^{(p+5)}\,,\,\,\bc^{(p+4)}\,,\,\, \bc^{(p+3)}\,,\,\, \bc^{(p+2)}&\cr O_{p-1}:&\bc^{(p+4)}\,,\,\,\bc^{(p+3)}\,,\,\, \bc^{(p+2)}\,,\,\, \bc^{(p+1)}& } \right. \nn\\
   F^{(p+4)}\,\rightarrow &&\,\left\{\matrix{O_p\,\,\,\,\,:&\bc^{(p+3)}\,,\,\,\bc^{(p+2)}\,,\,\, \bc^{(p+1)}\,,\,\, \bc^{(p)}&\cr O_{p-1}:&\bc^{(p+2)}\,,\,\,\bc^{(p+1)}\,,\,\, \bc^{(p)}\,,\,\, \bc^{(p-1)}&} \right. \nn\\
     F^{(p+2)}\,\rightarrow &&\,\left\{\matrix{O_p\,\,\,\,\,:&\bc^{(p+1)}\,,\,\,\bc^{(p)}\,,\,\, \bc^{(p-1)}\,,\,\, \bc^{(p-2)}&\cr O_{p-1}:&\bc^{(p)}\,,\,\,\bc^{(p-1)}\,,\,\, \bc^{(p-2)}\,,\,\, \bc^{(p-3)}&} \right. \nn\\
      F^{(p)}\,\rightarrow &&\,\left\{\matrix{O_p\,\,\,\,\,:&\bc^{(p-1)}\,,\,\,\bc^{(p-2)}\,,\,\, \bc^{(p-3)}\,,\,\, \bc^{(p-4)}&\cr O_{p-1}:&\bc^{(p-2)}\,,\,\,\bc^{(p-3)}\,,\,\, \bc^{(p-4)}\,,\,\, \bc^{(p-5)}&} \right. \nn\\
       F^{(p-2)}\,\rightarrow &&\,\left\{\matrix{O_p\,\,\,\,\,:&\bc^{(p-3)}\,,\,\,\bc^{(p-4)}\,,\,\, \bc^{(p-5)}\,,\,\, \bc^{(p-6)}&\cr O_{p-1}:&\bc^{(p-4)}\,,\,\,\bc^{(p-5)}\,,\,\, \bc^{(p-6)}\,,\,\, \bc^{(p-7)}&}  \right.\nn\\
 F^{(p-4)}\,\rightarrow &&\,\left\{\matrix{O_p\,\,\,\,\,:&\bc^{(p-5)}\,,\,\,\bc^{(p-6)}\,,\,\, \bc^{(p-7)}\,,\,\, \bc^{(p-8)}&\cr O_{p-1}:&\bc^{(p-6)}\,,\,\,\bc^{(p-7)}\,,\,\, \bc^{(p-8)}\,,\,\, \bc^{(p-9)}&} \right. \labell{FOO}
\eeqa
We have to impose the T-duality constraint \reef{OpOp} for each potential $\bc^{(n)}$. 

Let us begin with the most simple case. It can easily be observed  that the T-duality constraint  fixes the coefficient  of the coupling  $F^{(p-4)}$ to be zero. We  look at the term in the reduction which produces $\bc^{(p-9)}$. This term is produced only by the reduction of the coupling \reef{LP-4} when one of the transverse indices of the R-R field strength carries the $y$-index. The reduction of this term, however, is zero after imposing the O-plane conditions.  So this can not  constraint the coefficient of the coupling \reef{LP-4}. We   consider instead the reductions which produce $\bc^{(p-8)}$.  When the O$_p$-plane is along the circle, it produces the following reduction:
\beqa
\epsilon^{a_0\cdots a_{p-1}}\Big[\frac{a}{(p-4)!}\bH_{[ia_6a_7}\bc^{(p-8)}_{a_8\cdots a_{p-1}]} \bH^i{}_{a_0a_1}\bH_{ja_2a_3}\bH^j{}_{a_4a_5}\Big]+\cdots\labell{LP-4W}
\eeqa
where dots represent some other terms which do not include $\bc^{(p-8)}$. On the other hand, when O$_{p-1}$-plane is orthogonal to the circle, the reduction of the coupling \reef{LP-4} produces the following terms:
\beqa
\epsilon^{a_0\cdots a_{p-1}}\Bigg[\frac{a}{(p-4)!}\bH_{[ia_6a_7}\bc^{(p-8)}_{a_8\cdots a_{p-1}]} \bH^i{}_{a_0a_1}\Big(\bH_{ja_2a_3}\bH^j{}_{a_4a_5}+W_{a_2a_3}W_{a_4a_5}\Big)\Bigg]+\cdots\labell{LP-4T}
\eeqa
where dots represent some  terms with other structures.  The difference between this term and the T-duality transformation of  \reef{LP-4W} produces the following term which involves  $\bc^{(p-8)}$:
\beqa
\Delta^{p-8}&=&\epsilon^{a_0\cdots a_{p-1}}\Bigg[\frac{a}{(p-4)!}\bH_{[ia_6a_7}\bc^{(p-8)}_{a_8\cdots a_{p-1}]} \bH^i{}_{a_0a_1} W_{a_2a_3}W_{a_4a_5}\Bigg] \labell{Dp-8}
\eeqa
This term can not be cancelled by total derivative terms, so the T-duality constraint predicts the coefficient of the coupling \reef{LP-4} to be zero, \ie
\beqa
a=0\labell{a}
\eeqa
Hence, the T-duality constraint force the coupling \reef{LP-4} to be zero. It is a nontrivial result which  would be very difficult to confirm   with the S-matrix element of one R-R and three NS-NS vertex operators.

It can be also easily observed  that the T-duality constraint  fixes the coefficients  of the  $F^{(p+6)}$- couplings to be zero.  In this case we  look at the term in the reduction which produces $\bc^{(p+5)}$. This term is  produced only by the world-volume reduction of the couplings in  \reef{LP+6}.  The   T-duality transformation of this term    produces the following term for  $\bc^{(p+5)}$:
\beqa
\Delta^{p+5}&=&\epsilon^{a_0\cdots a_{p-1}}\frac{(-1)^pe^{\varphi}}{2p!}(3f_1-f_2) \bar{F}^{(p+6)}_{ijklmna_0\cdots a_{p-1}}W^{i}{}_o\bH^{ojk}\bH^{lmn}
\eeqa
which can not be cancelled by a gauge invariant total derivative term. Hence, the T-duality constraint  forces the above term to be zero, \ie $3f_1-f_2=0$.  To fix these coefficients completely, we  look also at the terms in the reduction which produce $\bc^{(p+4)}$.  The difference between the O$_{p-1}$-plane and the T-duality of O$_p$-plane produces many terms involving $\bc^{(p+4)}$. Here we focus on   the terms involving $\bc^{(p+4)}$ and  $\nabla\varphi$.  One can easily find that  only the  reduction of the second term in \reef{LP+6} produces   such term.  The T-duality of the reduction of  O$_p$-plane produces $\bar{F}^{p+5}\nabla\varphi\bH\bH$, whereas,  the reduction of  O$_{p-1}$-plane produces $\bar{F}^{p+5}\nabla\varphi WW$. They  can not cancel each other unless the coefficient of the second term in \reef{LP+6} to be zero, \ie $f_2=0$. Combining with the previous constraint, one finds
  \beqa
  f_1=0,\,f_2=0\labell{f12}
  \eeqa
   This is the result that the S-matrix calculation produces \cite{Garousi:2011ut}.

Since the coefficient of the $F^{(p-4)}$-coupling  is zero, the next simple case  to look at is  the terms involving $\bc^{(p-7)} $. One finds $\bc^{(p-7)} $ is produced only by the transverse reduction of  the couplings $F^{(p-2)}$ in \reef{Lp-2} which have R-R field strength with transverse indices.  Since only the couplings with coefficients $b_1,b_7$ in \reef{Lp-2} involves the R-R field strength with the transverse indices, and the transverse reduction of these terms produces non-zero results which are not total derivative terms, one finds that the T-duality constraint  \reef{OpOp}  fixes these coefficients to be zero, \ie 
\beqa
b_1=0,\,b_7=0\labell{b17}
\eeqa
 The above result can also be found by looking at the terms involving  $\bc^{(p-6)} $. One  finds that   only the reductions of the terms with coefficients $b_1,b_7$ survived the O-plane conditions. The T-duality constraint then  forces these coefficients to be zero.   This result is consistent with the S-matrix calculation \reef{Sb}.

The surviving terms  in \reef{Lp-2} have R-R field strength with  only world-volume indices. One finds that the reduction of these terms produce   terms involving  $\bc^{(p-5)} $. However, they are removed by the O-plane conditions. Having no $\bc^{(p-5)} $-term from the reduction of $ F^{(p-2)}$-couplings, one concludes that  the transverse reduction of $F^{(p)}$-couplings on the  O$_{p-1}$-plane which also produces $\bc^{(p-5)} $, must be zero.  So one has to consider the R-R field strengths $F^{(p)},\, \nabla F^{(p)}$ in \reef{LP} which have transverse indices because only those terms  produce $\bc^{(p-5)} $. In fact all terms in \reef{LP} have such structure. However, the transverse reduction of those terms that have only one transverse index, produce $\bH\wedge \bc^{(p-5)}$ with only world-volume indices which is removed by the O-plane condition.   Therefore,  they  produce no non-zero term after reduction. The terms in \reef{LP} which have more than one transverse indices, \ie $c_{21},c_{23},c_{34},c_{40}$, however, produce non-zero result  after imposing the O-plane conditions. The T-duality constraint \reef{OpOp} then requires these terms to be zero, \ie
\beqa
c_{21}=0,\, c_{23}=0,\,c_{34}=0,\,c_{40}=0\labell{c2123}
\eeqa
Since the reduced couplings involve only $\bc^{(p-5)}$ there is no total derivative terms connecting the reduced couplings. Moreover, since they involve no derivative of field strength $\bH$, there is no Bianchi identity relation between the reduced couplings.  Hence, the coefficients of all terms must be zero, as we have set in above equation. 

Since the coefficients of the couplings involving  $F^{(p+6)}$ are  zero,  \ie \reef{f12}, the next   simple case  to consider   is  to look at  the terms involving $\bc^{(p+3)} $.  One finds $\bc^{(p+3)} $ is produced only by the world-volume  reduction of  the couplings   in \reef{Lp+4} which have R-R field strength with no $y$ index. So all terms in  \reef{Lp+4}, except the terms  in which the R-R field strength carries the world-volume indices $a_0, \cdots, a_p$, produce $\bc^{(p+3)} $. The T-duality constraint \reef{OpOp} makes the coefficients of all these terms to be zero, \ie
\beqa
  e_{13}=0,\,e_{26}=0,\,e_{32}=0,\,e_{44}=0\labell{ee0}
 \eeqa
In finding the above result, we have  added all possible total derivative terms  and imposed the Bianchi identities and the $\epsilon$-tensor identities. We find that there is no total derivative term involved here.

There are still further T-duality constraint on the non-zero couplings involving $F^{(p+4)}$.  The T-duality constraint \reef{OpOp} produces  the following relations for the other coefficients: 
\beqa
&& e_{17}=0,\,e_{31}=0,\,e_{33}=0,\,e_{35}=0,\,e_{47}=0,\, e_6=0,\,e_9=0,\nn\\
&&e_{3}=e_1,\,e_{37}=-3e_1,\,e_{42}=-6e_1,\,e_{8}=-3e_1,\,e_{12}=-\frac{1}{2}e_1,\,e_{20}=\frac{3}{2}e_1,\,e_{28}=\frac{1}{2}e_1\labell{e8e37}
\eeqa
  In this case we find that there is some total derivative term involved in which we are not interested in this paper. Up to an overall coefficient $e_1$, then all terms in \reef{Lp+4} are fixed by the T-duality constraint that we have considered so far. 

It is interesting that the coefficients $e_8,\, e_{37}$ are identical which is in accord with the proposal that the second derivative of dilaton appears in the world-volume action as  the dilaton-Riemann curvature \reef{cR}. Moreover, the first derivative of dilaton appears only in the term with coefficient $e_3$. Using an integration by part on the first  term in  \reef{Lp+4}, and the relation $e_3=e_1$, one finds  that the first derivative of dilaton appears in the following extension of $\nabla_a\nabla^a H^{ABC}$:
\beqa
\nabla_a\nabla^a H^{ABC}&\rightarrow &\cD_a\nabla^aH^{ABC}\,\,\,\,;\,\,\, \cD_a\equiv \nabla_a-\nabla_a\Phi \labell{DDH}
\eeqa
We will see  that this structure appears in all couplings that the T-duality produces. Note that the transverse contraction of two derivatives, \ie $\nabla_i\nabla^i$ has been removed at the onset by  imposing the equations of motion.

Imposing  the constraints that we have found so far, \ie \reef{a}, \reef{f12}, \reef{c2123}, \reef{ee0}, and \reef{e8e37},  the remaining  reductions in \reef{FOO} are 
\beqa
   F^{(p+4)}\,\rightarrow &&\,\left\{\matrix{O_p\,\,\,\,\,:&\qquad\qquad \bc^{(p+1)}\,,\,\,\,\,\,\,\,\,\, \bc^{(p)}&\cr O_{p-1}:&\quad\qquad\bc^{(p+1)}\,,\,\, \bc^{(p)}\,\,\,,\,\,\,\, \bc^{(p-1)}&} \right. \nn\\
     F^{(p+2)}\,\rightarrow &&\,\left\{\matrix{O_p\,\,:&\bc^{(p+1)}\,,\,\,\bc^{(p)}\,\,\,,\,\, \bc^{(p-1)}\,,\,\, \bc^{(p-2)}&\cr O_{p-1}:&\bc^{(p)}\,,\,\,\bc^{(p-1)}\,,\,\, \bc^{(p-2)}\,\,\,,\,\,\,\,\, \bc^{(p-3)}&}  \right. \nn\\
      F^{(p)}\,\rightarrow &&\,\left\{\matrix{O_p\,\,\,\,\,:&\bc^{(p-1)}\,,\,\,\bc^{(p-2)}\,,\,\, \bc^{(p-3)}\,,\,\, \bc^{(p-4)}&\cr O_{p-1}:&\bc^{(p-2)}\,,\,\,\bc^{(p-3)}\,,\,\, \bc^{(p-4)}\,\,\,  \qquad \,\,\,\,&}   \right.\nn\\
       F^{(p-2)}\,\rightarrow &&\,\left\{\matrix{O_p\,\,\,\,\,:&\bc^{(p-3)}\,,\,\,\bc^{(p-4)}\,\,\,  \,\,\, &\cr O_{p-1}:&\bc^{(p-4)}\,\,\qquad\,\,\,\,\, \,\,\,  \,\,\,  &}  \right.  \labell{FOO1}
\eeqa
The next case that we are going to consider in the reductions \reef{FOO1}, is $ \bc^{(p+1)}$.  Since one part of the reduction involve the $F^{(p+2)}$-couplings, the T-duality constraint should relate the remaining constant $e_1$ in $F^{(p+4)}$-couplings to the $d$-parameters in \reef{Lp+2}. The T-duality constraint \reef{OpOp} in this case remarkably fixes $e_1$ and all $d$'s in terms of one overall parameter, \ie
\beqa
&&d_9=0,\,d_{10}=0,\,d_{22}=0,\,d_{36}=0,\,d_{41}=0,\,d_{42}=0,\,d_{43}=0,\,d_{47}=0,\nn\\
&&e_1=\frac{1}{12}d_{11},\,d_{12}=d_{11},\,d_{15}=-d_{11},\,d_{16}=-d_{11},\,d_2=\frac{1}{8}d_{11},\,d_{21}=-\frac{1}{4}d_{11},\nn\\
&&d_{26}=-\frac{1}{4}d_{11},\,d_{27}=-\frac{1}{8}d_{11},\,d_{29}=-\frac{1}{2}d_{11},\,d_3=-\frac{3}{8}d_{11},\,d_{30}=-\frac{1}{8}d_{11},\,d_{48}=\frac{1}{4}d_{11}\labell{dd}
\eeqa
 In this case also, the T-duality constraint requires some total derivative terms in which we are not interested. 

 The coefficients $d_{12},d_{15}$ in \reef{dd} are consistent with the S-matrix result \reef{d1112}. Moreover, the relation between $e_1$ and $d_{11}$ is also consistent with the S-matrix results \reef{d1112} and \reef{ee}.  As pointed out before, since $d_{11}=d_{12}$ the second derivative of dilaton appears as the dilaton-Riemann curvature \reef{cR}. The first derivative of dilaton also appears as dilaton-derivative extension of world-volume derivative contraction with Riemann curvature and with $H$, \ie
\beqa
\nabla_a R^{aABC}&\rightarrow &\cD_aR^{aABC}\nn\\
\nabla_a H^{aAB}&\rightarrow &\cD_aH^{aAB}\labell{DH}
\eeqa
Note that the transverse derivative contraction with the Riemann curvature and with $H$ have been removed by the equations of motion. We will see that this extension appears in other couplings that the T-duality  produces.

Since all  $e$-parameters and $d$-parameters are fixed up to the overall factor $d_{11}$, one does  not need to consider $\bc^{(p)}$ because this term is produced only by $F^{(p+4)}$- and $F^{(p+2)}$-couplings. In fact, we have checked that the T-duality constraint on $\bc^{(p)}$    reproduces only the   relations  in \reef{e8e37} and  \reef{dd}.
Hence, for the next case we consider $\bc^{(p-1)}$ in the reductions \reef{FOO1}. The T-duality constraint on this term should give some relations between  $F^{(p+4)}$-, $F^{(p+2)}$- and $F^{(p)}$-couplings. Since the parameters in the first two set of couplings are fixed, this constraint should fix the $c$-parameters in \reef{LP}. The T-duality constraint \reef{OpOp} in this case   fixes $d_{11}$ and  all $c$'s in terms of one overall parameter $c_{12}$, \ie
\beqa
&\!\!\!\!\!\!\!\!\!\!&c_{17}=0,\,c_{32}=0,\,c_{37}=0,\,c_{10}=0,\,c_{14}=0,\,c_{16}=0,\,c_{28}=0,\, c_{43}=0,\,c_7=0,\labell{cc}\\
&\!\!\!\!\!\!\!\!\!\!&d_{11}=2c_{12},\,c_{13}=\frac{1}{2}c_{12},\,c_2=-2c_{12},\,c_3=\frac{1}{2}c_{12},\,c_{33}=-\frac{1}{2}c_{12},\,c_{38}=\frac{1}{2}c_{12},\,c_{39}=-c_{12},\,c_{44}=2c_{12},\nn\\
&\!\!\!\!\!\!\!\!\!\!&c_{46}=-c_{12},\,c_5=-2c_{12},c_8=\frac{1}{2}c_{12},\,c_{24}=\frac{1}{4}c_{12},\,c_{30}=-\frac{1}{32}c_{12},c_{31}=\frac{1}{8}c_{12},c_{35}=-c_{12}\nn
\eeqa
  In this case also there are some total derivative terms in which we are not interested in this paper because we assumed the spacetime manifold has no boundary.

   The coefficients $c_2,c_3$ in \reef{cc} are consistent with the S-matrix result \reef{c2c3}. Moreover, the relation between $d_{11}$ and $c_2$ is also consistent with the S-matrix results \reef{c2c3} and \reef{d1112}. The coefficients $c_{12},\, c_{46}$ are not identical, so one may conclude that the corresponding  couplings in \reef{LP} are not  in accord with the proposal that the second derivative of dilaton appears in the world-volume action  as the dilaton-Riemann curvature \reef{cR}. However, using the R-R Bianchi identity \reef{Bian}, one can write
\beqa
\nabla_i F^{(p)}_{a_1\cdots a_p}=p \nabla_{a_1}F^{(p)}_{ia_2\cdots a_p}-\frac{p(p-1)}{2} H_{ia_1a_2}F^{(p-2)}_{a_3\cdots a_p}
\eeqa
where we have used the O-plane conditions  on $H$ and the fact that there is an overall tensor $\epsilon^{a_0\cdots a_p}$. Then   up to a total derivative term, one can write the term in \reef{LP} with coefficient $c_5$ as 
\beqa
\frac{1}{p!}\nabla_i F^{(p)}_{a_1\cdots a_p}H^{ia}{}_{a_0}\nabla_a\Phi &=&\frac{1}{(p-1)!}F^{(p)}_{ia_2\cdots a_p}H^{ia}{}_{a_1}\nabla_a\nabla_{a_0}\Phi+\frac{1}{(p-1)!}F^{(p)}_{ia_2\cdots a_p}\nabla_{a_0}H^{ia}{}_{a_1}\nabla_a\Phi\nn\\
&&-\frac{1}{2(p-2)!}H_{ia_1a_2}F^{(p-2)}_{a_3\cdots a_p}H^{ia}{}_{a_0}\nabla_a\Phi\labell{bia}
\eeqa
The first term on the right hand side  then has the same structure as the term with coefficient $c_{12}$. Since $c_{12}+c_5=c_{46}$, one can write the corresponding couplings in \reef{LP} as the dilaton-Riemann curvature \reef{cR}. The second term on the right hand side   can be combined with the first term in  \reef{LP} to write them as dilaton-derivative combination \reef{DDH}. The last term   should be added to the $b_9$-coupling in \reef{Lp-2}.

The coefficients $c_3,c_8$ are identical, hence, the corresponding couplings can be combined as the dilaton-derivative \reef{DDH}. It seems, however,  that the second derivative of dilaton in the coupling with coefficient $c_{13}$ in  \reef{LP} can not be combined with any coupling with structure $FHR$ to be written as the dilaton-Riemann curvature. This steams from the fact that when we have written the independent couplings in  \reef{LP}, we had not paid attention on the proposal \reef{cR}. Now that we have found the couplings we may use appropriate $\epsilon$-tensor identities to write the couplings as the dilaton-Riemann curvature. In fact, writing the world-volume indices explicitly as $0,1,\cdots, p$, one can find the following identity:
\beqa
\frac{1}{2(p-2)!}\, F_{jaa_{3}...a_{p}}\, H_{ia_{0}a_{1}} R^{iaj}\,_{a_{2}}  -\frac{1}{(p-1)!}\, F_{ja_{2}...a_{p}}\, H_{iaa_{0}}\, R^{iaj}\,_{a_{1}}  =\frac{1}{2(p-1)!}\, F_{ja_{2}...a_{p}}\, H_{ia_{0}a_{1}} R^{iaj}\,_{a}\nn
\eeqa
Using this $\epsilon$-tensor identity, one finds that the couplings in  \reef{LP} with coefficients $c_{13},c_{38},c_{39}$ can be written as the dilaton-Riemann curvature \reef{cR}.

  The T-duality constraint \reef{OpOp} for $\bc^{(p-2)}$ should  reproduce only the   relations  in \reef{cc}. We have checked it explicitly.  
  
  Finally, to relate the constant $c_{12}$ to the $b$-parameters in \reef{Lp-2} and $\alpha$-parameters in \reef{Tf21}, one can consider the T-duality constraint on $\bc^{(p-3)}$ or $\bc^{(p-4)}$. We consider $\bc^{(p-3)}$ in the reductions \reef{FOO1}. The T-duality constraint on this term should give some relations between  $F^{(p+2)}$-, $F^{(p)}$- and $F^{(p-2)}$-couplings and the   couplings in  \reef{Tf21}. Since the parameters in the first two sets of couplings are  fixed, this constraint should fix the $b$-parameters in \reef{Lp-2},  $\alpha$-parameters in \reef{Tf21} and $c_{12}$ in terms of one overall parameter. The T-duality constraint in this case   produces the following relations: 
  \beqa
  \alpha_2=-\alpha_1,\,b_2=-2\alpha_1,\,b_4=2\alpha_1,\,b_5=-2\alpha_1,\,b_9=-2\alpha_1,\,c_{12}=4\alpha_1
  \eeqa 
   In this case also there are some total derivative terms in which we are not interested in this paper.   The first relation above  is consistent with CS coupling \reef{Tf2}. The coefficients $b_2,b_4,b_5$ are consistent with the S-matrix result \reef{Sb}. The coefficient   $b_9$  is consistent with the proposal that the first derivative of dilaton appears in the dilaton-derivative combination. To see this we note that the last term in \reef{bia} has the same structure as $b_9$-coupling. Hence, this structure has coefficient $b_9-c_5/2=2\alpha_1$ which is  minus    of $b_2$. As a result they can be combined into the dilaton-derivative combination \reef{DDH}. This ends our illustrations that the T-duality constraint \reef{OpOp} can fix all parameters of the minimal gauge invariant couplings that we have found in section 2 up to an overall factor.

\section{Discussion}

In this paper, imposing  only the gauge symmetry and the T-duality symmetry on the effective action of O$_p$-plane,  we have found  the   following couplings at order $\alpha'^2$: 
\beqa
S&=&-\frac{\alpha_1T_p\pi^2\alpha'^2}{24}\int d^{p+1}x\Bigg[{\cal L}_{CS}^{(p-3)}+{\cal L}^{(p-2)}+{\cal L}^{(p)}+{\cal L}^{(p+2)}+{\cal L}^{(p+4)}\Bigg]\labell{Gen1}
\eeqa
where $\alpha_1$ is an overall constant that can not be fixed by the T-duality constraint. The gauge invariant Lagrangians are the following:
\beqa
{\cal L}_{\rm CS}^{(p-3)}& = &\epsilon^{a_0\cdots a_p}\bigg[\frac{ 1}{(p-3)!}{ C}^{(p-3)}_{a_4\cdots a_{p}}R_{a_{0}a_{1}}{}^{ij}R_{a_{2}a_3\,ij}-\frac{1}{(p-3)!}{ C}^{(p-3)}_{a_4\cdots a_{p}}R_{a_{0}a_{1}}{}^{ab}R_{a_{2}a_3\,ab}\bigg]\nn\\\labell{Tf22}
{\cal L}^{(p-2)}&\!\!\!\!\!=\!\!\!\!\!\!&2\epsilon^{a_{0}...a_{p}}\Big[- \frac{1}{(p-2)!}\, F_{a_{3}...a_{p}}\, \cD^{a}H_{iaa_{0}}\, H^{i}\,_{a_{1}a_{2}}\nn\\
&& 
+\frac{1}{(p-2)!}\, F_{a_{3}...a_{p}}\, \nabla_{a_{0}}H_{iaa_{1}}\, H^{ia}\,_{a_{2}}-\frac{1}{(p-2)!}\, F_{a_{3}...a_{p}}\, \nabla_{a}H_{ia_{0}a_{1}}\, H^{ia}\,_{a_{2}} \Big]\nn\\\labell{fLp-2}
{\cal L}^{(p)}&\!\!\!\!\!=\!\!\!\!\!\!& 4\epsilon^{a_{0}...a_{p}}\Big[ 
\frac{2}{(p-1)!}\,F_{ia_{2}...a_{p}}\, \cD_a\nabla_{a_{0}}H^{ia}\,_{a_{1}} -\frac{1}{2(p-1)!}\, F_{ia_{2}...a_{p}}\,\cD_a\nabla^{a}H^{i}\,_{a_{0}a_{1}} \nn\\
&&
-\frac{1}{(p-1)!}\, F_{ia_{2}...a_{p}}\, H^{i}\,_{aa_{1}}\, \cR^a{}_{a_0}  +\frac{1}{2(p-1)!}\, F_{ja_{2}...a_{p}}\, H_{ia_{0}a_{1}}\cR^{ij}  \nn\\
&&  
+\frac{1}{4(p-1)!}\, F_{ka_{2}...a_{p}}\, H_{iaa_{0}}\, H^{ijk}\, H_{j}\,^{a}\,_{a_{1}}  
-\frac{1}{32(p-3)!}\, F_{iaba_{4}...a_{p}}\, H^{iab}\, H^{j}\,_{a_{0}a_{1}}\, H_{ja_{2}a_{3}}\nn\\
&& +\frac{1}{8(p-1)!}\, F_{la_{2}...a_{p}}\, H_{ia_{0}a_{1}}\, H^{i}\,_{jk}\, H^{jkl}
 -\frac{1}{2(p-1)!}\, F_{ia_{2}...a_{p}}\, H^{iab}\, H^{j}\,_{aa_{0}} H_{jba_{1}}
 \nn\\
&&-\frac{1 }{(p-1)!}\, F_{ka_{2}...a_{p}}\, H^{ijk}\, R_{ia_{0}ja_{1}}
+\frac{2}{(p-1)!}\, F_{ia_{2}...a_{p}}\, H^{iab}\, R_{aa_{0}ba_{1}} \Big]\nn\\\labell{fLP}
{\cal L}^{(p+2)}&\!\!\!\!\!=\!\!\!\!\!\!&8\epsilon^{a_{0}...a_{p}}\Big[ \frac{1}{8(p+1)!}\, \nabla_{k}F_{la_{0}...a_{p}}\, H^{ijk}\, H_{ij}\,^{l}-\frac{3}{8(p+1)!}\, \nabla_{i}F_{ja_{0}...a_{p}}\, H^{iac}\, H^{j}\,_{ac}
\nn\\
&& 
 +\frac{1}{p!}\, F_{ij a_{1}...a_{p}}\,\cD_a R^{iaj}\,_{a_{0}}
 -\frac{1}{4p!}\, F_{jk a_{1}...a_{p}}\,\cD^aH_{ia a_{0}}\, H^{ijk}  
 \nn\\
&& 
+\frac{1}{(p+1)!}\, \nabla_{i}F_{ja_{0}...a_{p}}\,\cR^{ij}  -\frac{1}{4p!}\, F_{jk a_{1}...a_{p}}\, \nabla^{a}H^{i jk}\, H_{ia a_{0}} -\frac{1}{8p!}\, F_{kl a_{1}...a_{p}}\, \nabla_{a_{0}}H^{ij k}\, H_{ij}\,^{l}
 \nn\\
&&-\frac{1}{2p!}\, F_{ij a_{1}...a_{p}}\, \nabla_{a}H^{i}\,_{ba_{0}}\, H^{jab}
-\frac{1}{8p!}\, F_{ij a_{1}...a_{p}}\, \nabla_{a_{0}}H^{i}\,_{ab}\, H^{jab}
\Big]\nn\\\labell{fLp+2}
{\cal L}^{(p+4)}&\!\!\!\!\!=\!\!\!\!\!\!&\frac{2}{3} \epsilon^{a_{0}...a_{p}}\Big[-\frac{1}{(p+1)!}F_{ijk a_{0}...a_{p}}\,\cD_a\nabla^{a}H^{ijk}-\frac{3}{(p+1)!}\, F_{jkla_{0}...a_{p}}\, H^{ikl}\,\cR^j{}_i  
 \nn\\
&&
 -\frac{1}{2(p+1)!}\,  F_{kmna_{0}...a_{p}} H^{ijk} H_{i}\,^{lm} H_{jl}\,^{n} +\frac{3}{2(p+1)!}\,  F_{jkla_{0}...a_{p}}\, H_{iab}\, H^{ikl}\, H^{jab}    \nn\\
&&  
+\frac{1}{2(p+1)!}\, F_{ijka_{0}...a_{p}}\, H^{iab}\, H^{j}\,_{ac}\, H^{k}\,_{b}\,^{c}  
-\frac{6}{(p+1)!}\,  F_{ijka_{0}...a_{p}}\, H^{i}\,_{ab}\, R^{ja kb}
 \Big]\labell{fLp+4}
\eeqa
The second derivative of dilaton appears in the dilaton-Riemann curvature \reef{cR} and the first derivative of dilaton appears in the dilaton-derivative \reef{DDH}.  
Most of the  couplings in \reef{Gen1} are new couplings which have not been found by any other method in string theory. This action   is  fully consistent with the partial couplings that have been already found  in the literature by the S-matrix method, \ie the couplings of one arbitrary R-R field strength and one NS-NS, and also the couplings of one R-R field strength $F^{(p-2)}$ and two B-fields.

We have seen that the O-plane couplings at order $\alpha'$, found by the T-duality constraint, are  the same as the orientifold projection of the partial couplings that have been found in the literature from the disk-level S-matrix elements. However, the world-sheet corresponding to the tree-level S-matrix elements of O-plane is $RP^2$. This may indicate that the orientifold projection of disk-level S-matrix elements and the $RP^2$-level S-matrix elements should have the same low energy expansion at order $\alpha'^2$. In other worlds,  up to overall factors,  the orientifold projection of D$_p$-brane couplings at order $\alpha'^2$ should produce  the O$_p$-plane couplings  at order $\alpha'^2$. This is not, however, the case for higher orders of $\alpha'$ which can be seen from the curvature expansion of the anomalous CS couplings, \ie
\beqa
\sqrt{{\cal L}(\pi^2\alpha'R)}&=&1+\frac{(4\pi^2\alpha')^2}{96}p_1(R)-(4\pi^2\alpha')^4\left(\frac{1}{10240}p_1^2(R)-\frac{7}{23040}p_2(R)\right)+\cdots\nn\\
\sqrt{{\cal A}(4\pi^2\alpha'R)}&=&1-\frac{(4\pi^2\alpha')^2}{48}p_1(R)+(4\pi^2\alpha')^4\left(\frac{1}{2560}p_1^2(R)-\frac{1}{2880}p_2(R)\right)+\cdots
\eeqa
where the first one is for O-plane and the second one is for D-brane \cite{Morales:1998ux}. The reason that  the couplings are proportional at order $\alpha'^2$ but not at the higher orders, may be rooted to the fact that the T-duality transformation at order $\alpha'^2$ has no higher derivative correction whereas one expects corrections to the Buscher rules at higher orders of $\alpha'$. 
If the T-duality transformations are the Buscher rules  \reef{T2} which are linear, then the T-duality constraint would satisfy at each order of $\alpha'$  separately. The resulting couplings at a given order of $\alpha'$ then can be divided to two parts by the orientifold projection. One part would be the O-plane couplings.  However, the corrections to the Buscher rules which are not linear,  mix  the constraints at different orders of $\alpha'$. That is, the constraints at a given order of $\alpha'$ has contribution from the couplings at that order as well as couplings at lower orders of $\alpha'$. Then the orientifold projection of the resulting T-duality invariant couplings at the given order of $\alpha'$ would not be the same as the couplings that one would find by imposing the orientifold projection at all orders of $\alpha'$. Hence, the orientifold projection of the D-brane couplings at  order  $\alpha'^3$ hand higher would not produce the  corresponding O-plane couplings.    

The disk-level S-matrix elements of one arbitrary R-R and two NS-NS vertex operators have been calculated in \cite{Velni:2013jha,Becker:2016bzb}. The low energy expansion of them should produce D-brane  couplings at order $\alpha'^2$. The orientifold projection of those couplings  should then be the same as the couplings that we have found in \reef{Gen1}.  It would be interesting to perform this calculation.

We have seen that the derivatives of dilaton appears  only  through the dilaton-Riemann curvature \reef{cR}  and the dilaton-derivative \reef{DDH}. It has been shown in \cite{Garousi:2010ki}  that the dilaton-Riemann curvature is invariant under  linear T-duality. The dilaton-derivative is also invariant under the linear T-duality. In fact one can write the contraction of the dilaton-derivative with an arbitrary vector at the linear order of metric perturbation as 
\beqa
\cD_aA^a&=&\prt_a A^a+\frac{1}{2}A^a\eta^{bc}\prt_a h_{bc}-\prt_a\Phi A^a
\eeqa
where $G_{AB}=\eta_{AB}+h_{AB}$. Separating the world-volume indices to $y$-index and other world-volume indices, and using the  linear T-duality transformations $h_{yy}\rightarrow -h_{yy}$ and $\Phi\rightarrow \Phi-h_{yy}/2$, then one finds the above expression is invariant under the linear T-duality. Similar analysis has been done in \cite{Garousi:2010ki} to show that the dilaton-Riemann curvature is invariant under the linear T-duality. The invariance of the world-volume action under  linear T-duality requires the derivatives of dilaton appear in the dilaton-Riemann and dilaton-derivative combinations. However, the invariance of the effective action under full nonlinear T-duality requires that the couplings of one R-R and  an arbitrary number of NS-NS fields appear only through the combination \reef{Gen1}.

The action \reef{Gen1} is complete action of O$_p$-plane at order $\alpha'^2$  for $\alpha_1=-1/4$. This action however has only  one R-R field. The O$_p$-plane action for zero R-R field have been found in \cite{Robbins:2014ara,Garousi:2014oya}. This  action should have  couplings involving two, three and four R-R fields as well. Each set of couplings may be found by the T-duality constraint up to an overall factor. Then the S-duality may be used to relate the overall factor of  three R-R couplings to  the couplings \reef{Gen1}, and the two and four R-R couplings to the couplings found in \cite{Robbins:2014ara,Garousi:2014oya}. It would be interesting to perform this calculation to find  a gauge invariant action which is also invariant under the T-duality and the S-duality. 

It would be also interesting to extend the calculation in this paper to find  the D$_p$-brane couplings at order $\alpha'^2$. A difficulty in this calculation is that  each coupling   in the effective action at order $\alpha'^2$ may have an arbitrary number of $B_{ab}$. They may also have  world-volume derivative of this field, \ie  $\prt_a B_{bc}$ which does not appear in the field strength $H_{abc}$. They are consistent with the gauge symmetry because the D-brane has also open string gauge field strength $f_{ab}$ and the combination $B_{ab}+f_{ab}$ is invariant under the gauge transformation. The T-duality does not relate the massless closed string fields to the massless open string fields. Hence, in the T-duality constraint for the massless closed string fields, one may have  couplings that are not gauge invariant. The reduction of those couplings then would not be invariant under the $U(1)\times U(1)$ gauge transformations. That makes problem in using the trick used  in section 3  to  keep only the  $U(1)\times U(1)$ gauge invariant part of reduction of the Riemann curvature and other field strengths.

In finding the parameters in section 4, we have ignored some total derivative terms in the base space. If O-plane are at the fixed point of closed spacetime, then there would be  no boundary in the base space and the total derivative terms become zero by using the Stokes's theorem. However, if the spacetime has boundary, then the base space has boundary as well. In this case, the O-plane may end to the boundary. Hence, the total derivative terms in the base space can not be ignored. They produce some boundary terms in the boundary of the base space \cite{Garousi:2019xlf}. In that case, one should consider some couplings at the boundary of O-plane.  The boundary terms in the boundary of the base space should be cancelled by the  T-duality of the couplings on the boundary of O-plane. This constraint may fix the couplings at the boundary of the O-plane.  It would be interesting to find the  boundary terms in the effective action of O-plane. 

 \vskip .3 cm
{  \bf Acknowledgments}:   This work is supported by Ferdowsi University of Mashhad under grant  3/45013(1396/08/02).


\end{document}